\documentclass[final,1p,10pt,a4paper]{elsarticle}
\pdfoutput=1

\usepackage{graphicx}
\usepackage{amssymb,amsmath}

\biboptions{sort&compress}

\newcommand{\Eq}[1]{\mbox{Eq.\,\eqref{#1}}}
\newcommand{\Fig}[1]{\mbox{Fig.\,\ref{#1}}}
\newcommand{\Tab}[1]{\mbox{Table\,\ref{#1}}}
\newcommand{\Sec}[1]{\mbox{Sec.\,\ref{#1}}}

\newcommand{\phys}{\mathrm{phys}}

\begin{document}

\newcommand{\preprintnumbers}{\setlength{\unitlength}{1mm}
\begin{picture}(0,0)
  \put(0,30){\scriptsize ADP-12-29/T796, DESY 12-105, Edinburgh 2012/11}
\end{picture}}

\begin{frontmatter}

\date{August 27, 2012}

\title{\preprintnumbers
 Nucleon mass and sigma term from lattice QCD with two light
 fermion flavors}

\author[ur]{G.S.~Bali}
\author[ur]{P.C.~Bruns}
\author[ur]{S.~Collins}
\author[ur]{M.~Deka}
\author[ur]{B.~Gl\"a\ss{}le}
\author[ur]{M.~G\"ockeler}
\author[ur]{L.~Greil}
\author[ur]{T.R.~Hemmert}
\author[ed]{R.~Horsley}
\author[ur]{J.~Najjar}
\author[RIKEN]{Y.~Nakamura}
\author[jsc]{A.~Nobile}
\author[ur,jsc]{D.~Pleiter}
\author[liv]{P.E.L.~Rakow}
\author[ur]{A.~Sch\"afer}
\author[ur]{R.~Schiel}
\author[desy]{G.~Schierholz}
\author[ur]{A.~Sternbeck\corref{cor}}\ead{andre.sternbeck@ur.de}
\author[adl]{J.~Zanotti}

\address{%
\begin{center}
 \small\textnormal{(QCDSF Collaboration)}
\end{center}
}

\cortext[cor]{Corresponding author}

\address[ur]{Institut f\"ur Theoretische Physik, Universit\"at Regensburg,  
        93040 Regensburg, Germany}

\address[ed]{School of Physics, University of Edinburgh, Edinburgh EH9
3JZ, UK}

\address[RIKEN]{RIKEN Advanced Institute for Computational Science, Kobe,
Hyogo 650-0047, Japan}

\address[jsc]{JSC, Research Center J\"ulich, 52425 J\"ulich, Germany}

\address[liv]{Theoretical Physics Division, Department of Mathematical
Sciences, University of Liverpool, Liverpool L69 3BX, UK}

\address[desy]{Deutsches Elektronen-Synchrotron (DESY), 22603 Hamburg,
Germany} 
\address[adl]{School of Chemistry and Physics, University of
  Adelaide, SA 5005, Australia}

\begin{abstract}
We analyze $N_f=2$ nucleon mass data with respect to their dependence
on the pion mass down to $m_{\pi}= 157$\,MeV and compare it with predictions
from covariant baryon chiral perturbation theory (BChPT). A novel
feature of our approach is that we fit the nucleon mass data
simultaneously with the directly obtained pion-nucleon
$\sigma$-term. Our lattice data below $m_{\pi}= 435$\,MeV is well described by
$O(p^4)$ BChPT and we find $\sigma=37(8)(6)$\,MeV for the
$\sigma$-term at the physical point. Using the nucleon mass to set the
scale we obtain a Sommer parameter of $r_0=0.501(10)(11)\,\text{fm}$.
\end{abstract}


\end{frontmatter}

\section{Introduction}

Predicting low-energy hadronic properties is one basic goal of lattice QCD.
A particular challenge to experiment as well as to theory is posed by the
so-called pion-nucleon $\sigma$-term
\begin{equation}
 \sigma = m_\ell \left\langle N | (\bar{u} u + \bar{d} d )| N \right\rangle
 \label{eq:sigmadef}
\end{equation}
which parametrizes the light quark contribution to the nucleon mass.
Here $m_\ell = m_u = m_d$ denotes the light quark mass.

At present, phenomenology does not give a clear picture of the magnitude
of $\sigma$. A dispersion theoretical analysis led to 
\mbox{$\sigma = 64(8)\,\text{MeV}$}~\cite{Koch:1982pu}, similar to the value
\mbox{$\sigma = 64(7)\,\text{MeV}$} obtained later in 
Ref.~\cite{Pavan:2001wz}. However the analysis of Ref.~\cite{Gasser:1990ce} 
suggested a much lower value, \mbox{$\sigma =
  45(8)\,\text{MeV}$}, which was also found in \cite{Borasoy:1996bx}. 
Recently, a new evaluation resulted in \mbox{$\sigma = 59(7)\,\text{MeV}$}
\cite{Alarcon:2011zs}.

Calculating the pion-nucleon $\sigma$-term directly on the lattice is
a computationally intensive task as it involves the computation of
quark-line disconnected correlation functions. This has become
feasible recently and was performed by us and
others~\cite{Babich:2010at,Bali:2011ks,Dinter:2012tt}, though only at
a single value of $m_\ell$ and of the lattice spacing $a$.
An alternative, which has often been used in the past (see, e.g.,
the recent studies \cite{Horsley:2011wr,Durr:2011mp,Dinter:2012tt,
Shanahan:2012wh}), is given by the Feynman-Hellmann theorem. It allows us to
express $\sigma$ in terms of the derivative of the nucleon mass $M_N$ with
respect to $m_\ell$: 
\begin{equation}
 \sigma = m_\ell \frac{\partial M_N}{\partial m_\ell} \,.
 \label{eq:sigmafh}
\end{equation}

Calculating the nucleon mass on the lattice as a function of $m_\ell$ is
relatively straightforward, but becomes expensive close to the physical
point. 

Until very recently, lattice QCD calculations have therefore been
performed at rather large quark masses, such that results had to be
extrapolated to the physical point over a wide range. Thanks to the
efforts of different lattice QCD collaborations during the last years
this situation has much improved. The QCDSF Collaboration, for
example, has generated a large set of $N_f=2$ gauge field
configurations for a variety of quark masses, lattice spacings and
volumes, reaching down to pion masses of about 157 MeV. These
simulations employ the standard Wilson gauge action and the
non-perturbatively $O(a)$ improved clover action for two flavors of
mass-degenerate quarks.

In this paper we analyze nucleon mass data obtained from these
simulations with respect to their quark-mass and volume dependence,
compare this to $SU(2)$ covariant baryon chiral perturbation theory
(BChPT), and extract a value for the pion-nucleon $\sigma$-term
utilizing the Feynman-Hellmann theorem.  A novel feature of our
analysis is that we combine the nucleon mass data with a direct
determination of the pion-nucleon \mbox{$\sigma$-term \cite{Bali:2011ks}},
fitting both \emph{simultaneously} to the corresponding $O(p^4)$ BChPT
expressions. It turns out that this gives much more reliable and
precise results compared to fits to the nucleon mass data only.

The outline of this article is as follows: In the next section we
summarize and describe our lattice data. Our fitting procedure is
described in \Sec{sec:fittingformulae}. The results of the fits are
then discussed in Sec.~\ref{sec:results}. In \Sec{sec:conclusion} we
summarize the outcome of our analysis. In \mbox{\ref{app:BChPT}} we review
the results from BChPT which underlie our analysis. Some more details
on the fits can be found in \ref{app:fitdetails}.

\section{Lattice data}
\label{sec:latticedata}

Our lattice data for the pseudoscalar ($am_\pi$) and nucleon mass
($aM_N$) were extracted from one-exponential fits to smeared-smeared
correlators. For the smearing we used Jacobi smearing for all data
sets apart from the more recent analyses at $\beta=5.29$,
$\kappa=0.13632$ and $\kappa=0.13640$.  For $\kappa=0.13632$ our
aforementioned direct calculation of the $\sigma$-term was performed
\cite{Bali:2011ks}, for which a more optimized smearing was critical
to obtaining a signal for the scalar matrix element. We achieved this
by using Wuppertal smearing with APE smoothed links.
This improved method was also applied to our calculations at
$\kappa=0.13640$.

\Tab{tab:data} lists our data for the pseudoscalar ($am_\pi$) and
nucleon masses ($aM_N$). We also state the corresponding values
$r_0m_\pi$ and $r_0M_N$ in units of the Sommer scale $r_0$,
extrapolated to $m_\ell=0$ \cite{BaliNajjar} (see
\Tab{tab:r0c_a_values}). The error quoted for $am_\pi$ and $aM_N$ is
the statistical uncertainty of the data, while for $r_0m_\pi$
($r_0M_N$) it is the combined error of $r_0/a$ and $am_\pi$ ($aM_N$).

\begin{table*}
\centering\small
\begin{tabular}{rcclllll}\hline\hline
  \multicolumn{1}{c}{\phantom{$\star$\;}$\beta$} & $\kappa$ & lattice & $am_\pi$
& $aM_N$ & $r_0m_\pi$ & $r_0M_N$
& $L/r_0$
\\\hline\\*[-2ex]
 5.25 & 0.13460 & $16^3\times32$ &   0.4932(10) &   0.9436(49) &   3.256(27) &   6.230(59) & 2.42\\
 5.25 & 0.13520 & $16^3\times32$ &   0.3821(13) &   0.7915(55) &   2.523(22) &   5.226(56) & 2.42\\
$\star$\; 5.25 & 0.13575 & $24^3\times48$ &   0.2556(5) &   0.6061(38) &   1.687(14) &   4.002(41) & 3.63\\
$\star$\; 5.25 & 0.13600 & $24^3\times48$ &   0.1840(7) &   0.5088(72) &   1.215(11) &   3.359(55) & 3.63\\
 5.25 & 0.13620 & $32^3\times64$ &   0.0997(11) &   0.4012(87) &   0.658(9) &  
2.649(61) & 4.85\\*[0.5ex]
 5.29 & 0.13400 & $16^3\times32$ &   0.5767(11) &   1.0546(51) &   4.039(32) &   7.386(67) & 2.28\\
 5.29 & 0.13500 & $16^3\times32$ &   0.4206(9) &   0.8333(32) &   2.946(24) &   5.836(50) & 2.28\\
 5.29 & 0.13550 & $12^3\times32$ &   0.3605(32) &   0.8325(96) &   2.525(30) &   5.831(81) & 1.71\\
 5.29 & 0.13550 & $16^3\times32$ &   0.3325(14) &   0.7020(72) &   2.329(20) &   4.916(63) & 2.28\\
$\star$\; 5.29 & 0.13550 & $24^3\times48$ &   0.3270(6) &   0.6858(33) &   2.290(18) &   4.804(44) & 3.43\\
 5.29 & 0.13590 & $12^3\times32$ &   0.3369(62) &   0.8071(208) &   2.360(47) &   5.653(152) & 1.71\\
 5.29 & 0.13590 & $16^3\times32$ &   0.2518(15) &   0.6306(53) &   1.763(17) &   4.417(50) & 2.28\\
$\star$\; 5.29 & 0.13590 & $24^3\times48$ &   0.2395(5) &   0.5554(46) &   1.677(13) &   3.890(44) & 3.43\\
$\star$\; 5.29 & 0.13620 & $24^3\times48$ &   0.1552(6) &   0.4670(49) &   1.087(10) &   3.271(42) & 3.43\\
 5.29 & 0.13632 & $24^3\times48$ &   0.1112(9) &   0.4250(60) &   0.779(9) &   2.977(48) & 3.43\\
 5.29 & 0.13632 & $32^3\times64$ &   0.1070(5) &   0.3900(50) &   0.750(7) &   2.732(41) & 4.57\\
$\star$\; 5.29 & 0.13632 & $40^3\times64$ &   0.1050(3) &   0.3810(30) &   0.735(6) &   2.669(29) & 5.71\\
 5.29 & 0.13640 & $40^3\times64$ &   0.0660(8) &   0.3708(196) &   0.463(7) &   2.597(139) & 5.71\\
$\star$\;  5.29 & 0.13640 & $48^3\times64$ &   0.0570(7) &
 0.3420(80) &   0.399(6) &   2.395(59) & 6.85\\*[0.5ex]
 5.40 & 0.13500 & $24^3\times48$ &   0.4030(4) &   0.7556(17) &   3.339(30) &   6.260(58) & 2.90\\
 5.40 & 0.13560 & $24^3\times48$ &   0.3123(7) &   0.6260(26) &   2.588(24) &   5.186(51) & 2.90\\
 5.40 & 0.13610 & $24^3\times48$ &   0.2208(7) &   0.5085(45) &   1.829(17) &   4.213(53) & 2.90\\
 5.40 & 0.13625 & $24^3\times48$ &   0.1902(6) &   0.4655(35) &   1.576(15) &   3.857(45) & 2.90\\
 5.40 & 0.13640 & $24^3\times48$ &   0.1538(10) &   0.4265(67) &   1.274(14) &   3.533(64) & 2.90\\
$\star$\; 5.40 & 0.13640 & $32^3\times64$ &   0.1505(5) &   0.4163(44) &   1.246(12) &   3.449(48) & 3.86\\
 5.40 & 0.13660 & $32^3\times64$ &   0.0845(6) &   0.3530(71) &   0.700(8) &   2.924(64) & 3.86\\
$\star$\; 5.40 & 0.13660 & $48^3\times64$ &   0.0797(3) &   0.3143(52) &   0.660(7) &   2.604(49) & 5.79\\
\hline\hline
\end{tabular} 
\caption{Lattice data for the pseudoscalar ($am_\pi$) and the nucleon
mass ($aM_N$) in lattice units.  In columns 6 and 7 we list the
corresponding values in units of $r_0$. For $am_\pi$ and $aM_N$ we
give the statistical errors, for $r_0m_\pi$ ($r_0M_N$) the errors are
the combined statistical errors of $r_0/a$ and $am_\pi$ ($aM_N$).
Stars in the first column mark all those entries which enter our fits
[depending on the upper limit set for $(r_0m_\pi)^2$].}
\label{tab:data}
\end{table*} 

\begin{table}[b]
\centering
 \begin{tabular}{c@{\qquad}c@{\qquad}c@{\qquad}c}
 \hline\hline
  \rule{0ex}{2.2ex} $\beta$ & 5.25 & 5.29 & 5.40 \\
 \hline
    \rule{0ex}{2.2ex} $r_0/a$ & 6.603(53) & 7.004(54) & 8.285(74)\\
 \hline\hline
 \end{tabular} 
 \label{tab:r0c_a_values}
\caption{Lattice estimates of $r_0/a$ for different $\beta$, extrapolated
to the chiral limit \cite{BaliNajjar}.}
\end{table} 

\begin{figure}[t]
\centering
\includegraphics[width=0.9\textwidth]{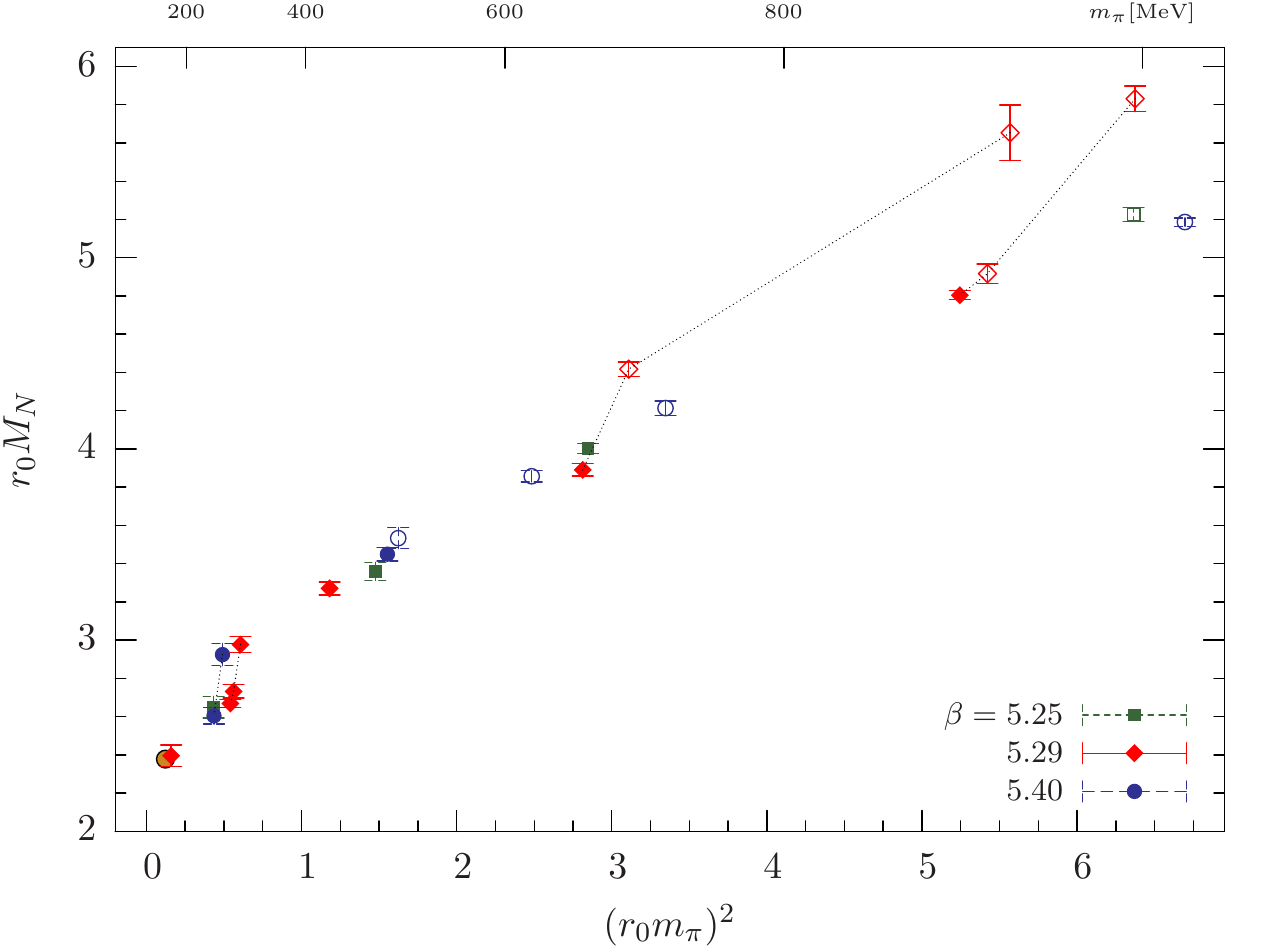}
\caption{Lattice data for $r_0M_N$ versus $(r_0m_\pi)^2$ for different
lattice spacings ($\beta=5.25$, $5.29$, $5.40$) and volumes. Open
(filled) symbols refer to points where $L\le3r_0$ ($L>3r_0$). A
black-framed (yellow) circle indicates the physical point assuming
$r_0=0.5$\,fm. The dotted lines connect points of the same
$(\beta,\kappa)$ but different lattice size. They are meant to guide
the eye and to illustrate finite-volume effects.}
\label{fig:raw_data}
\end{figure}

In Fig.~\ref{fig:raw_data} we show our data for $r_0M_N$ plotted
versus $(r_0m_\pi)^2$. Different symbols and colors are used to
distinguish between data for the different $\beta$. Data points which
refer to the same $(\beta,\kappa)$ but different lattice volumes are
connected by dotted lines to emphasize finite volume effects.

From this figure (and also from closer inspection of the data) we find
that, within the given precision, our data show no systematic
dependence on the lattice spacing and, thus, are close to the
continuum limit. We therefore scale our data by the respective
values of the chirally extrapolated $r_0/a$ and do not attempt a
continuum-limit extrapolation. Finite volume effects, on the other
hand, are clearly visible and will be incorporated into the fits. In
fact, they provide an additional valuable input, because some of
the low-energy constants (LECs) also enter the volume corrections.

\begin{figure}[t]
  \centering
 \mbox{\includegraphics[width=0.45\linewidth]{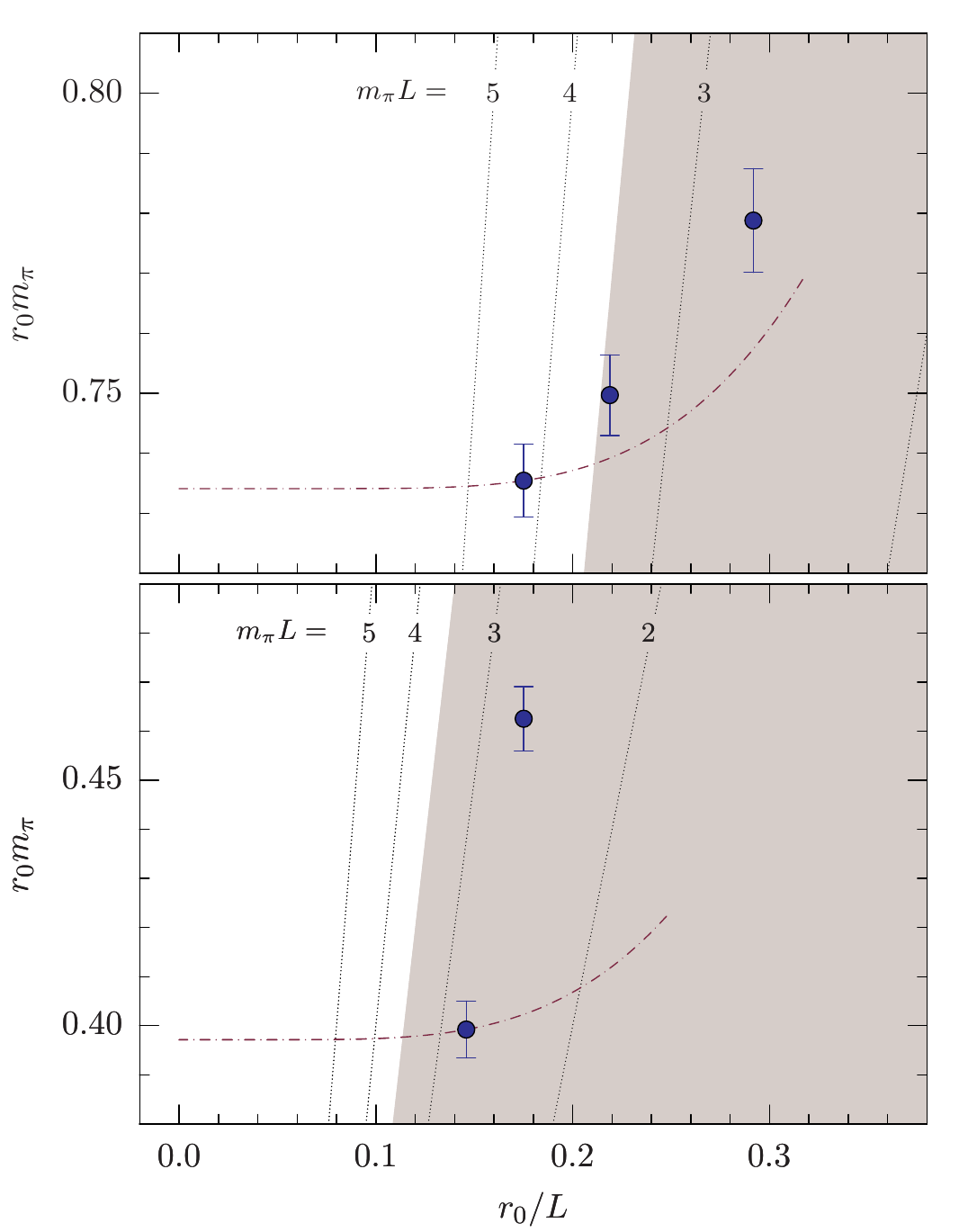}
\qquad
 \includegraphics[width=0.45\linewidth]{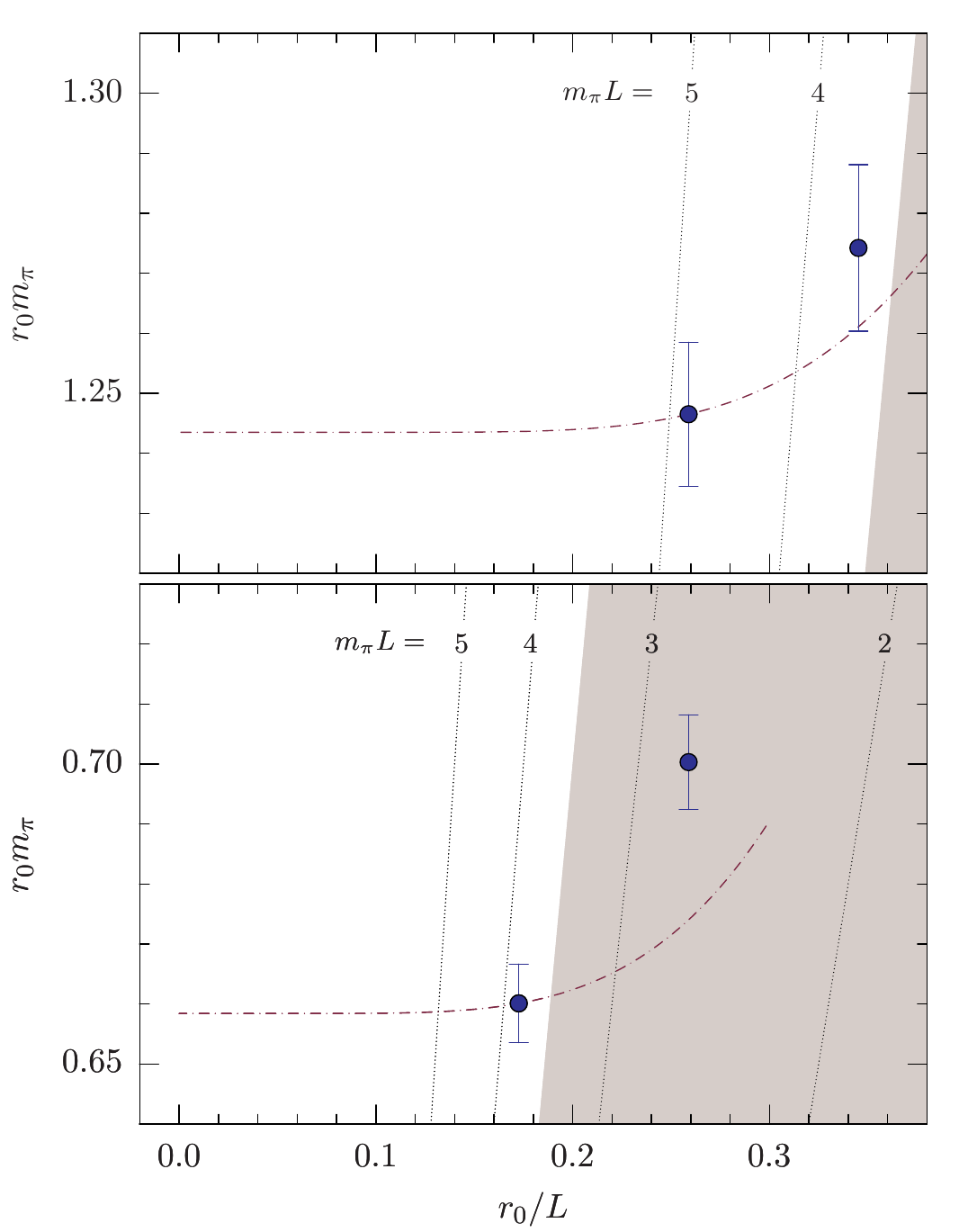}}
  \caption{Volume dependence of our pion mass data. Left top: $\beta=5.29$,
    $\kappa=0.13632$. Left bottom: $\beta=5.29$,
    $\kappa=0.13640$. Right top: $\beta=5.40$, $\kappa=0.13640$. Right
    bottom: $\beta=5.40$, $\kappa=0.13660$. Dashed-dotted
    lines represent finite-volume extrapolations, obtained from
    applying the method of \cite{Colangelo:2005gd} to the point for
    the smallest $r_0/L$. Vertical dotted lines denote
    constant $m_\pi L$. Gray areas mark the region ($m_\pi L<3.5$) where
    finite-volume corrections are not under control.}
  \label{fig:r0mpi_L_r0_extrapolation}
\end{figure}

Before fitting these data to BChPT expressions, all values for the
pion mass have to be extrapolated to infinite volume. A method to
calculate these finite-volume corrections has been worked out in
\cite{Colangelo:2005gd}. When applying this method to our data we
find, however, that the corresponding values for $m_\pi L$ have to
satisfy at least $m_\pi L>3.5$. Below this limit, the finite-size
effects of our pion mass data are stronger than the calculated
corrections. If $m_\pi L\ge 4$ the data points and the extrapolated
values even agree within errors (see
\Fig{fig:r0mpi_L_r0_extrapolation} for an illustration). This means
that, within the available precision, correcting for finite-volume
effects according to \cite{Colangelo:2005gd} will only help us if
$3.5\le m_\pi L\le4.0$. In all these cases we have at least one point
with $m_\pi L\approx4$ (or larger).

For the final fits to BChPT, we therefore exclude all ($\beta$, $\kappa$)
combinations for which there is not at least one data point that satisfies
$m_\pi L>3.5$, and then take the value for $r_0m_\pi$ from the largest available
volume (with $m_\pi L>4$) as our estimate of the
infinite-volume limit. As argued above, a finite-volume correction
according to Ref.~\cite{Colangelo:2005gd} would not give different
results for $r_0m_\pi$, but in this way we are a bit more conservative
about the error. 

There is, however, one data point where we allow for an exception to
this rule: the estimate for $r_0m_\pi$ at $\beta=5.29$ and
$\kappa=0.13640$ ($m_\pi\approx157\,\textrm{MeV}$). At these parameters
the largest available lattice size is currently $48^3\times64$, but we
expect no drastic changes of $r_0m_\pi$ if the volume is enlarged
further, because the PCAC-mass dependence of our pion mass data below
290 MeV (considering largest-volume data only) is
quite linear already.%
\footnote{Note also that if with larger volumes the pion mass
  further decreased the $\chi^2_r$-value of our fits presented
  below would actually improve, with negligible effects on the fitted
  parameters. We have tested that.} 

\medskip
As mentioned above, a novel feature of our analysis is that we take
additional data for $\sigma$ into account, which comes
from a direct determination. So far, we have computed this at
$\beta = 5.29$, $\kappa = 0.13632$ on a $32^3\times64$ and $40^3 \times 64$
lattice \cite{Bali:2011ks}. This corresponds to a pion mass of about 
290 MeV, and we will use the value
\begin{equation}
  r_0\sigma = 0.273(25)
  \label{eq:r0sigma_meas}
\end{equation}
from the $40^3 \times 64$ lattice in what follows.

\section{Fitting to covariant chiral perturbation theory}
\label{sec:fittingformulae}

\subsection{Fitting formulae}

We intend to fit our data to BChPT. The expressions at
next-to-leading one-loop order are given in \ref{app:BChPT}. 
We expand those up to order $m_\pi^4$:
\begin{align}
  M_N = M_0 &- 4c_1 m_\pi^2 
                    - \frac{3g_A^2 m_\pi^3}{32\pi F_\pi^2}
                    + 4e_1^r m_\pi^4 \nonumber\\
                    &+ \frac{m_\pi^4}{8\pi^2 F_\pi^2} 
                   \left[\frac{3c_2}{16}
                    - \frac{3g_A^2}{8M_0}
                    + \log\frac{m_\pi}{\lambda}
                    \left(8c_1 - \frac{3c_2}{4} - 3c_3 - \frac{3g_A^2}{4M_0}
                    \right)\right]\,,
\label{eq:Mn_exp}
\end{align} 
\begin{align}
 \sigma = -&4c_1 m_\pi^2 - \frac{9g_A^2 m_\pi^3}{64\pi F_\pi^2}
             + m_\pi^4 \left[8e_1^r -\frac{8c_1 l_3^r}{F_\pi^2}
                        +\frac{3c_1}{8\pi^2 F_\pi^2}
                        -\frac{3c_3}{16\pi^2 F_\pi^2}\right.\nonumber\\
                                &\left.
                        -\frac{9g_A^2}{64\pi^2 M_0 F_\pi^2} + \frac{1}{4\pi^2
F_\pi^2}                                  \log\frac{m_\pi}{\lambda}
                                  \left( 7c_1 - \frac{3c_2}{4}
                                     -3c_3 - \frac{3g_A^2}{4M_0}\right)
                                \right]\;.
\label{eq:sigma_exp}
\end{align}
These expressions involve the low-energy constants $c_1$, $c_2$ and
$c_3$ as well as the renormalized counterterm coefficient $e_1^r$ and
\begin{equation}
 l_3^r\equiv-\frac{1}{64\pi^2}
          \left(\bar{l}_3 + 2\log\frac{m^{\mathrm{phys}}_\pi}{\lambda}\right)\,,
\end{equation}
which depend on the renormalization scale $\lambda$. The pion decay
constant $F_\pi$ and the nucleon axial coupling constant $g_A$ are
taken at the physical point, which is consistent with the order of
BChPT we are using. For the fits we will set $F_\pi=92.4$\,MeV and
$g_A=1.256$. The renormalization scale $\lambda$ is set to 
$\lambda=m_\pi^{\mathrm{phys}}=138\,\text{MeV}$.
\footnote{Note that
  choosing another renormalization scale will only change the values of
  $e^r_1$ and $l_3^r$. All other parameters are invariant. We
  have checked that this is satisfied for our fits.}   

To correct for finite-volume effects in our nucleon mass data we employ the
finite-volume correction at next-to-leading one-loop order in BChPT.
It reads
\begin{equation}
  \Delta M_N(m_\pi^2,L)=\Delta M^{(3)}(m_\pi^2,L)+\Delta
  M^{(4)}(m_\pi^2,L) \,,
  \label{eq:DeltaMn}
\end{equation}
where $\Delta M^{(3)}$ and $\Delta M^{(4)}$ can be read off
from Eqs.~\eqref{eq:deltam3} and \eqref{eq:deltam4} with the
substitution $\overline{m}\to m_\pi$. This is consistent with
the order we are using.

Since our data is given in units of $r_0$, for the fits we have to
re-express all dimensionful quantities in units of $r_0$, too: 
\begin{equation}
 M_N(m_\pi)\to \widehat{M}_N(\widehat{m}_\pi),\quad
\Delta M_N(m_\pi,L)\to \Delta\widehat{M}_N(\widehat{m}_\pi,\widehat{L})
\quad\text{and}\quad
\sigma(m_\pi)\to\widehat{\sigma}(\widehat{m}_\pi)\;
\end{equation}
where a ``$\widehat{\phantom{...}}$'' indicates the expression is understood in
units of $r_0$, e.g., $\widehat{M}_N = r_0 M_N$.

As not all parameters of Eqs.~\eqref{eq:Mn_exp} and
\eqref{eq:sigma_exp} are well constrained by our fits, some of these
will be fixed to their phenomenological values, e.g., $F_\pi$, $c_2$,
$c_3$ and $\bar{l}_3$. Consequently, a \emph{physical} input value
for $r_0$ has to be provided as well, which is however unknown a
priori.

We fix $r_0$ for each fit separately by iterating over
different physical $r^{(k)}_0$ ($k=1,2,\ldots$), plugged into the
fitting formulae until $r_0$ comes out self-consistently from the fit,
that is 
\begin{equation}
 \label{eq:r0iteration}
 \epsilon > \left\vert r_0^{(k)} - r_0^{(k-1)}\right\vert\quad\text{where}\quad
 r_0^{(k)} = \frac{\widehat{M}_N\left(r_0^{(k-1)}\cdot
m^{\mathrm{phys}}_\pi\right)}{
M_N^{\mathrm{phys}}}\:.
\end{equation}
We set $\epsilon=0.001$, and for most of our fits $r_0$ converges
after 10 to 15 iterations. By construction, all fits will pass through
the physical point $M^{\mathrm{phys}}_N=938\,\text{MeV}$ at
$m^{\mathrm{phys}}_\pi=138\,\text{MeV}$.

Two kinds of fits will be discussed below: fits to
our nucleon mass data, using the $\chi^2$-function
\begin{equation}
 \chi^2_N = \sum^{N_{\mathrm{data}}}_{i=1}\frac{\left(\widehat{M}_N(x_i) +
\Delta \widehat{M}_N(x_i,y_i) - z_i\right)^2}{e^2_i}\;
\label{eq:chi2Mn}
\end{equation}
and combined (simultaneous) fits to the nucleon and
$\sigma$-term data, using the $\chi^2$-function
\begin{equation}
 \chi^2_{N\!\sigma} =  \chi^2_N + \frac{\left(\widehat{\sigma}_N(x_j) -
\bar{z}_j\right)^2}{\bar{e}^2_j}\;.
\label{eq:chi2Mnsigma}
\end{equation}
Here $x=r_0m_\pi$, $y=L/r_0$, $z=r_0M_N$ refer to our measured points and $e$
denotes the error of $r_0M_N$.
In $\chi^2_{N\!\sigma}$, $\bar{z}_j$ ($j\in \{1, \ldots , N_{\mathrm{data}} \}$)
refers to our single directly determined result for $r_0\sigma$ 
and $\bar{e}_j$ is its error, see \Eq{eq:r0sigma_meas}. As finite-size
effects appear to be negligible for this number (see Ref.~\cite{Bali:2011ks})
we do not apply any finite-volume corrections in this case.

\subsection{Fit ranges}

Our nucleon mass data of Table~\ref{tab:data} covers a range of
$r_0m_\pi$ values from 0.42 up to 4.04, and $L/r_0$ ranges from 1.71
to 6.85. In physical units this corresponds to \mbox{$m_\pi\approx
0.17\ldots1.58\,\text{GeV}$} and $L=0.85\ldots 3.4\,\text{fm}$ when
$r_0=0.5\,\text{fm}$. When fitting these data we do not know a priori
for which range of values of $m_\pi$ and $L$ we can trust our fitting
functions. We therefore vary constraints on $r_0m_\pi$ ($L/r_0$) from
above (below).  The constraint $L/r_0>3$ turns out to be low enough
such that there is sufficient data to perform stable fits but also
large enough so that $\Delta M_N$ captures the finite-volume
effects, see, e.g., the discussion below.

For $r_0m_\pi$, on the other hand, we find it reasonable to constrain
it from above by $(r_0m_\pi)^2_{\max}=1.6$, if not
$(r_0m_\pi)^2_{\max}=1.3$.  Both these upper bounds give results which
agree within errors, albeit with a larger uncertainty for the
latter. Also our independent measurement of $r_0\sigma$
[\Eq{eq:r0sigma_meas}] at $r_0m_\pi\approx 0.735$ is then well
reproduced by \Eq{eq:sigma_exp}, not only for a combined fit
[\Eq{eq:chi2Mnsigma}] but also if one uses \Eq{eq:sigma_exp} with the
parameters from a stand-alone fit to the nucleon mass data.  For
larger $(r_0m_\pi)^2_{\max}$, say $(r_0m_\pi)^2_{\max}=3.0$, the fits
change quantitatively and qualitatively: The $m_\pi$-dependence of the
nucleon mass [\Eq{eq:Mn_exp}] becomes more concave in shape and our
data point for $r_0\sigma$ lies below the fit curves. If one only
considered the nucleon mass data, such fits over a larger
$m_\pi$-range would roughly capture the overall $m_\pi$-dependence
[even up to $m_\pi\approx 1$~GeV where $O(p^4)$ BChPT certainly does
not hold], but this is likely to be accidental for the given order as
was pointed out already in \cite{McGovern:2006fm}. Our definitive
analysis will therefore be restricted to fits for which
$(r_0m_\pi)^2_{\max}\le1.6$. For our final estimates of $r_0$ and
$\sigma$ we will even restrict ourselves to
$(r_0m_\pi)^2_{\max}\le1.3$.

\subsection{Parameters}
\label{sec:parameters}

Let us now discuss the parameters in the fit functions. Besides $F_\pi$ and
$g_A$, the functions in Eqs.~\eqref{eq:chi2Mn} and \eqref{eq:chi2Mnsigma}
contain five and six free parameters, respectively. These are $M_0$,
$c_i$ ($i=1,2,3$) and $e^r_1$ for \Eq{eq:chi2Mn} and $\bar{l}_3$ in addition 
for \Eq{eq:chi2Mnsigma}. Ideally one would like to determine these all from fits
to the lattice data. This is however not possible with our current data
and we therefore have to fix some of the parameters to values from the
literature.

As $\bar{l}_3$ only enters $r_0\sigma$ [see \Eq{eq:sigma_exp}] this
parameter should not be left free in a fit with only one data 
point to constrain it. For our combined fits [\Eq{eq:chi2Mnsigma}] we therefore
fix it to the FLAG-estimate \cite{Colangelo:2010et}
\begin{equation}
 \bar{l}_3=3.2(8)
 \label{eq:l3bar}
\end{equation} 
and check for the stability of our final results varying $\bar{l}_3=3.2$ within
one standard deviation.

For $c_2$ and $c_3$ we proceed similarly. At first one might be tempted not to
fix these two parameters at all. However, like the renormalization-scale
dependent parameter $e^r_1$, $c_2$ and $c_3$ gain influence at larger
$m_\pi$. So, if these parameters are all left free in fits to 
low $m_\pi$-mass data, their uncertainties would be unreasonably
large. We therefore decided to fix $c_2$ and $c_3$ to their latest
phenomenological values 
\cite{Meissner:2005ba, Meissner:2011ba},
\begin{equation}
   c_2 = 3.3(2)\,\textrm{GeV}^{-1} \quad\text{and}\quad c_3
=-4.7(1.3)\,\textrm{GeV}^{-1} \,,
 \label{eq:c2c3_Meissner}
\end{equation} 
and investigate the stability of our final results varying the less precisely
known parameter $c_3$ by one standard deviation. Additional fits are performed
where $c_3$ is left as a free parameter. 

To check the stability of the fit parameters $c_1$ and $M_0$, and also of our
estimates of $r_0M_N^{\mathrm{phys}}$ and $r_0\sigma$ at the physical
point, additional fits are performed where the expressions in
Eqs.\eqref{eq:Mn_exp} and \eqref{eq:sigma_exp} are truncated at orders
$O(m_\pi^3)$ and $O(m_\pi^2)$.

\section{Results}
\label{sec:results}

\subsection{Discussion of our fits}

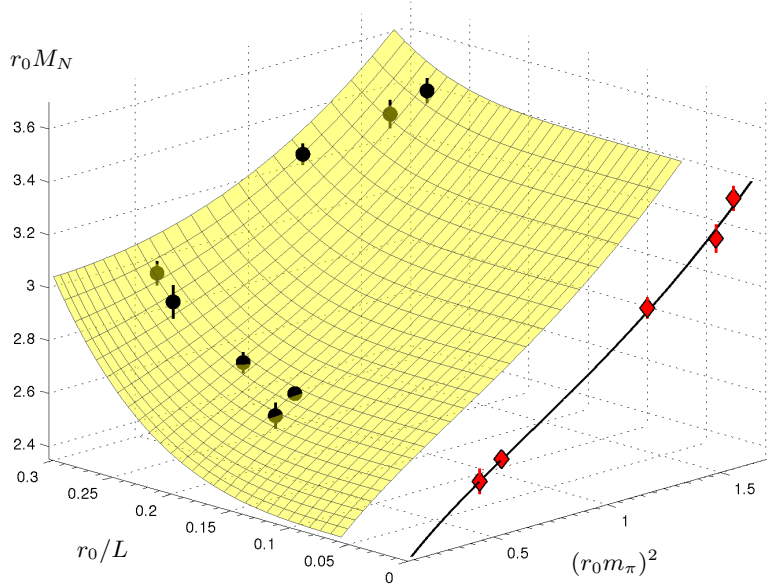
\begin{figure*}
 \centering
  \begin{picture}(340,220)
\put(0,0){%
\includegraphics[width=0.9\linewidth]{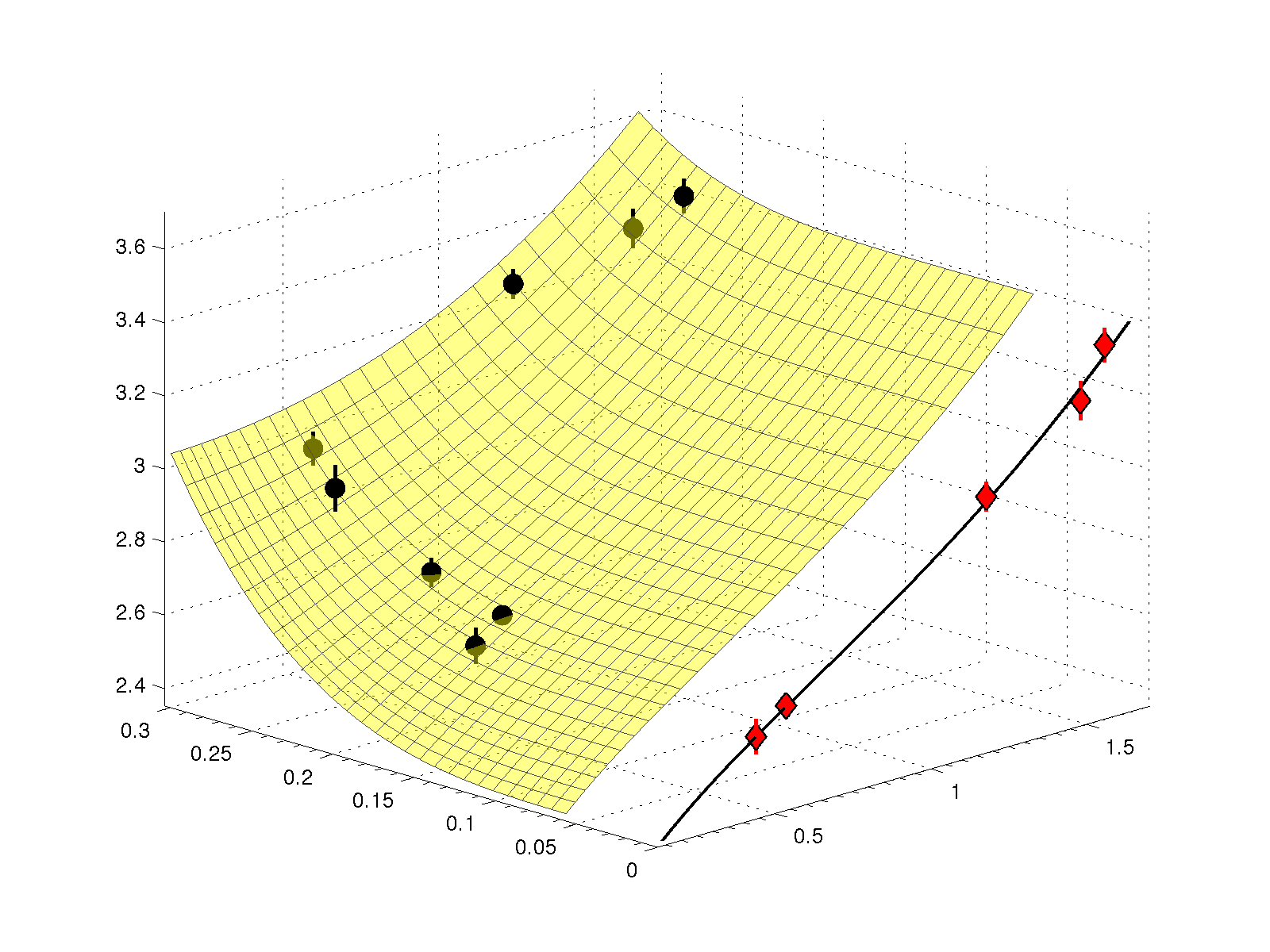}}
\put(30,215){\small $r_0M_N$}
\put(55,30){\small $r_0/L$}
\put(240,25){\small $(r_0m_\pi)^2$}
\end{picture}
 \caption{Fit (surface) to the nucleon mass data (full circles) for a
   range of $r_0/L$ and $(r_0m_\pi)^2$ values. Black circles lie above
   the surface, gray (or half gray) below (or on) the surface. The line
   and the red diamonds at $r_0/L=0$ mark the fitted infinite volume
   prediction.  For the sake of simplicity, for each $(r_0m_\pi)^2$ only
   one extrapolated (red) point is shown at $r_0/L=0$.}
 \label{fig:3dPlot}
\end{figure*} 

In \Fig{fig:3dPlot} we show an example fit to the nucleon mass
data. Full (black) circles in \Fig{fig:3dPlot} represent our lattice
data for the nucleon mass, and the surface is a fit to these
points. We fit to the surface $r_0M_N(r_0m_\pi,L/r_0)$, hence the 
finite-volume corrections are determined directly through the fit as
well. The (red) diamonds in \Fig{fig:3dPlot} represent the lattice
data, after subtracting these volume corrections. 

An overview of our fits and parameters can be found in
\ref{app:fitdetails} where we list all our combined and stand-alone
fits in Tables \ref{tab:fitparaS} and \ref{tab:fitparaN},
respectively, and also provide more detail for the interested reader. 

\begin{figure}
  \centering
  \mbox{\includegraphics[height=12.8cm]{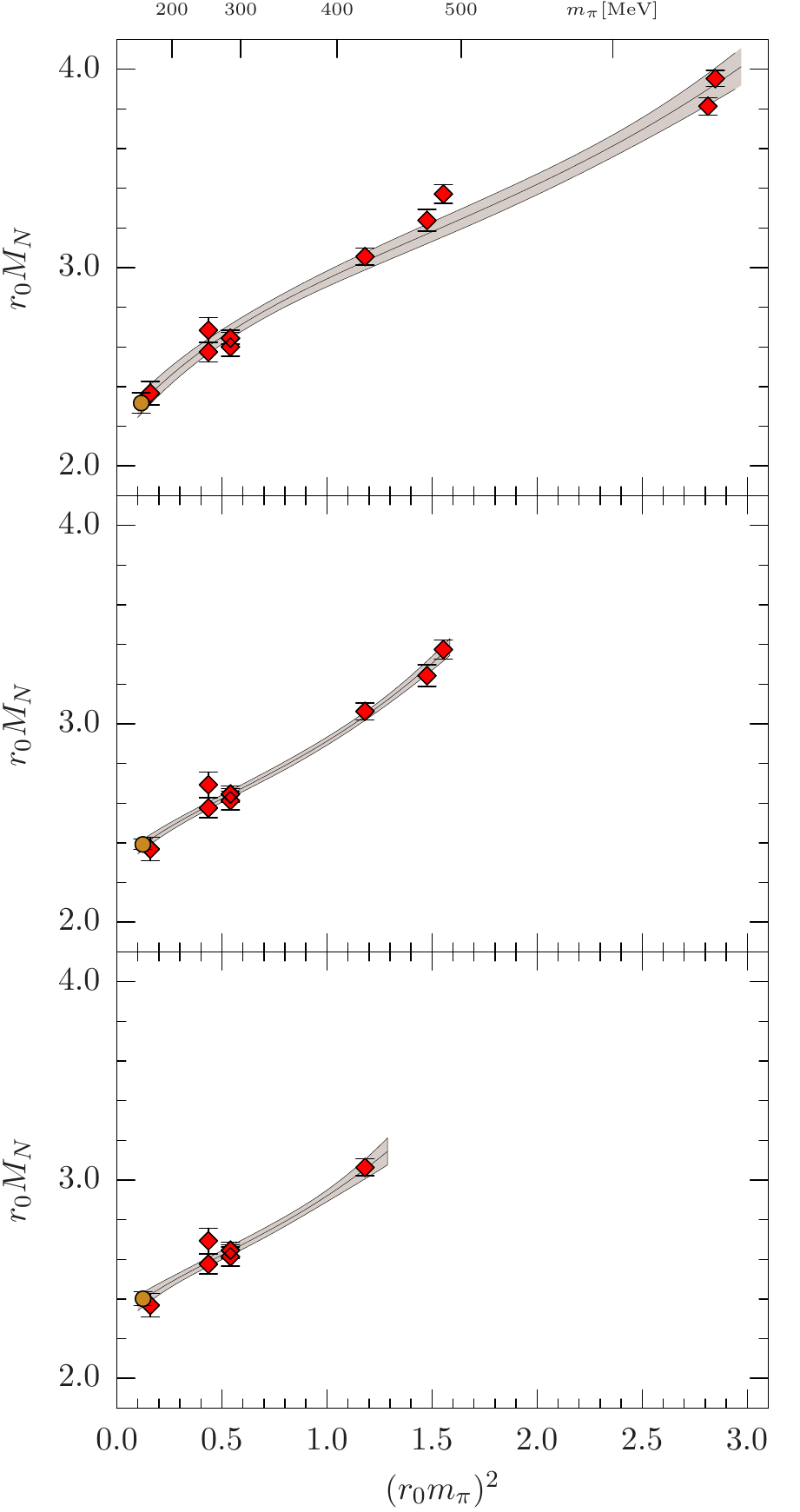}
  \includegraphics[height=12.8cm]{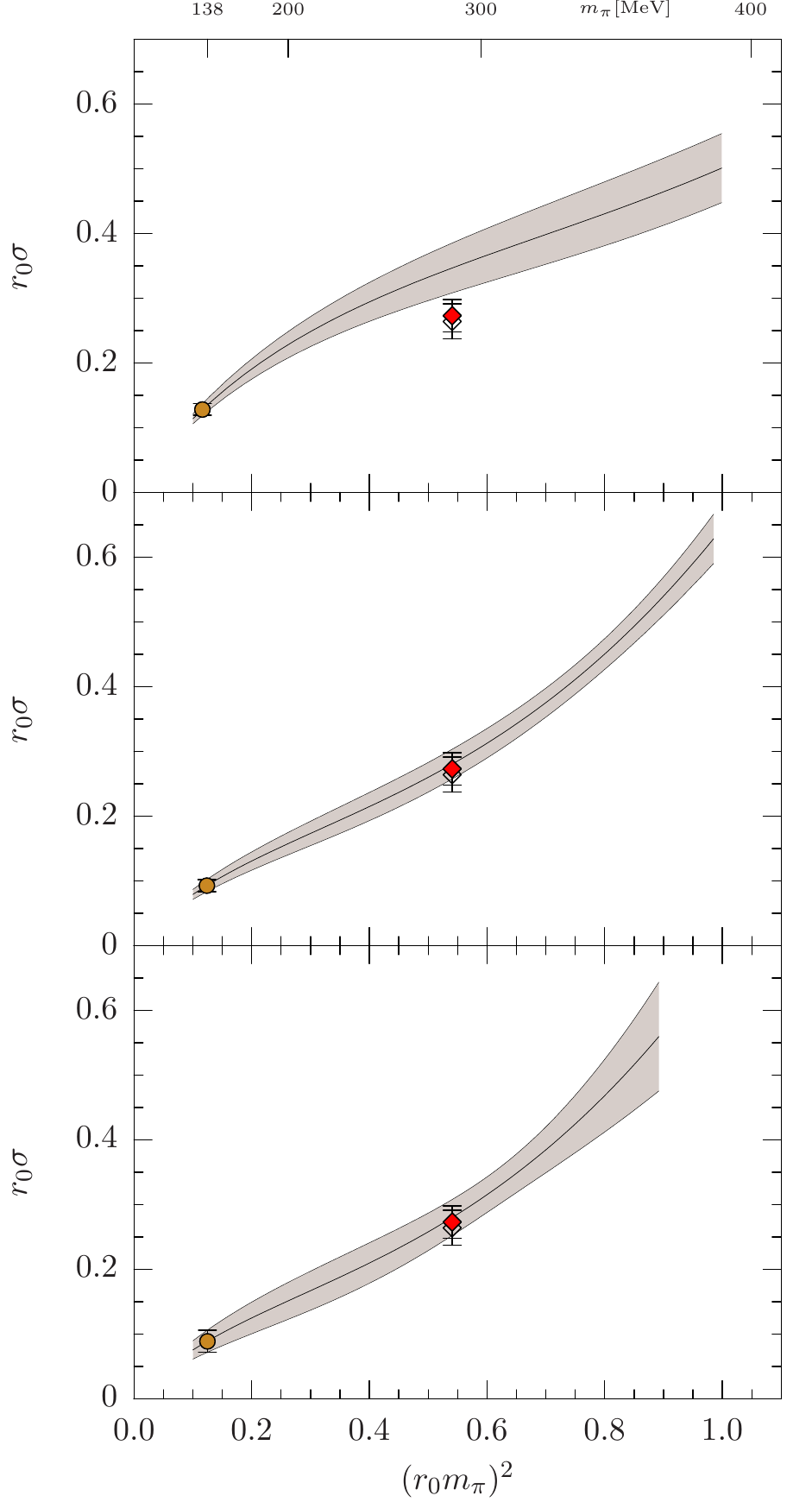}}
 \caption{Simultaneous fits to the nucleon mass (left) and $\sigma$-term
  data (right) for three fitting windows with $(r_0m_\pi)_{\max}^2=3.0$, 1.6 
  and 1.3 (from top to bottom). These fits are labeled \texttt{Soo3},
  \texttt{Soo2} and \texttt{Soo1} in \Tab{tab:fitparaS}, where
  $c_2\equiv 3.3\,\mathrm{GeV}^{-1}$, $c_3\equiv -4.7\,\mathrm{GeV}^{-1}$ and
  $\bar{l}_3\equiv3.2$. Lines, error bands and 
  (full red) points are shown for the limit $L\to\infty$ (cf.~\Fig{fig:3dPlot}).
  A black-framed circle marks the location of the physical point using---for
  each plot separately---the $r_0$-value for which the fit is
  self-consistent. Open points did not enter any fit. In the left
  plots, the overlap of red points at $(r_0m_\pi)^2=0.436$ and 0.538
  indicates the quality of the (fitted) finite-volume corrections.}
 \label{fig:Splots}
\end{figure}

\begin{figure}
  \centering
 \mbox{\includegraphics[width=0.45\linewidth]{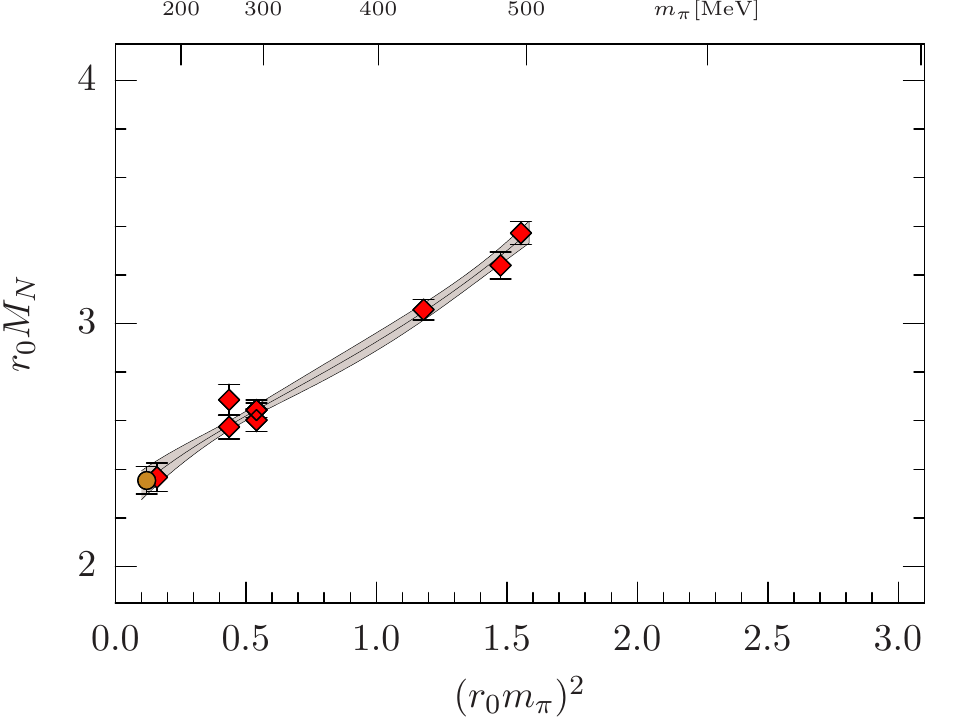}
 \includegraphics[width=0.45\linewidth]{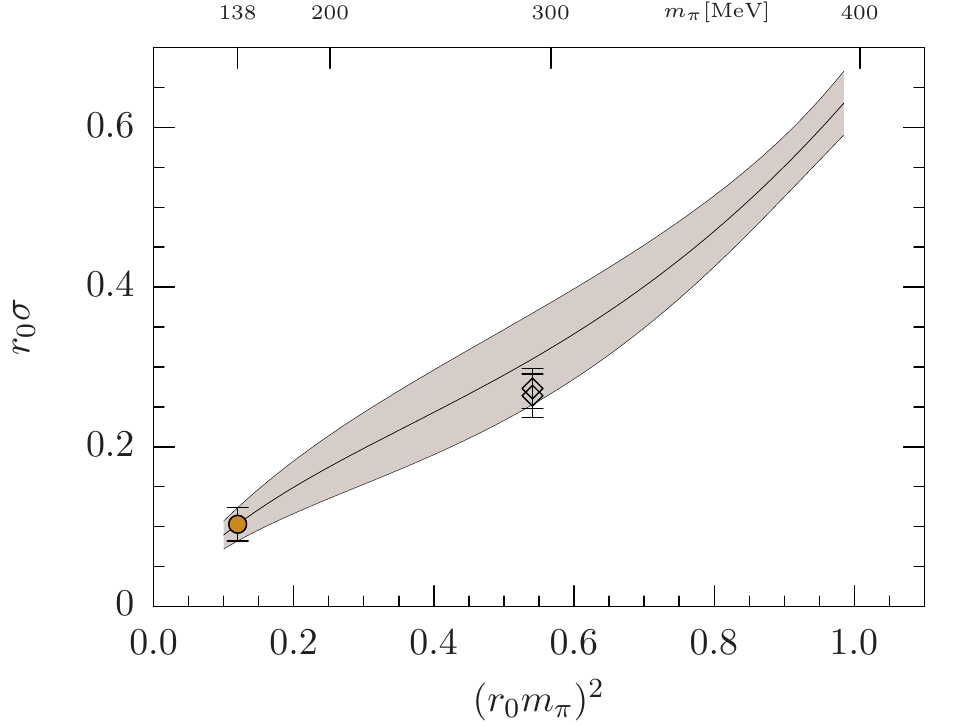}}
 \caption{Left panel: stand-alone fit to the nucleon mass data 
  (see fit \texttt{Noo2} in \Tab{tab:fitparaN}). 
  Right panel: comparison of the $\sigma$-term data
  (open symbols) for a $32^3\times 64$ and  $40^3\times64$ lattice
 \cite{Bali:2011ks} and the BChPT expression for
  $\sigma(m_\pi)$ with the parameters $M_0$, $c_1$ and $e_r^1$ taken from
  the nucleon mass fit shown in the left panel. The black-framed circle is the
  value for $r_0M_N$ ($r_0\sigma_{N}$) at the  physical point for these
  parameters. In both panels, the one-sigma error band is shown in gray.
  As in \Fig{fig:Splots}, the overlap of (red) points (left panel) at same
  $(r_0m_\pi)^2$ indicates the quality of the (fitted) finite-volume
  corrections.}
 \label{fig:Nplot}
\end{figure}

In Figs.~\ref{fig:Splots} and \ref{fig:Nplot} we display the
fit curves corresponding to some of the results listed in these
tables, together with the lattice data after subtracting the volume
corrections. The overlap of points indicates the quality of these
fitted corrections. In \Fig{fig:Splots}, three of our combined fits are shown
(\texttt{Soo3}, \texttt{Soo2} and \texttt{Soo1} of \Tab{tab:fitparaS}),
each for the same choice of fixed parameters ($c_2$, $c_3$
and $\bar{l}_3$ as given in Eqs.~\eqref{eq:l3bar}
and \eqref{eq:c2c3_Meissner}), but for different fit ranges
\begin{equation}
  (r_0m_\pi)^2<(r_0m_\pi)^2_{\max} \,,
\end{equation}
where $(r_0m_\pi)^2_{\max}$ is $3.0$, $1.6$ or $1.3$, always requiring
$L/r_0 > 3$. \Fig{fig:Nplot} shows a corresponding stand-alone fit
(\texttt{Noo2} of \Tab{tab:fitparaN}) to the nucleon mass data 
for $(r_0m_\pi)^2_{\max}=1.6$.

Our fits perform significantly better if the $\sigma$-term
constraint [\Eq{eq:r0sigma_meas}] is incorporated. For these combined
fits the individual uncertainty of each fit parameter 
is smaller than for the corresponding fit to the nucleon mass
data alone, while the $\chi^2_r$-values \mbox{($\chi^2_r\equiv\chi^2/ndf$)}
largely remain unaffected. Fits qualities are inferior, if nucleon
mass data up to $(r_0m_\pi)^2 = 3.0$ is included, but if one lowers this 
upper bound to 1.6 or 1.3, $\chi^2_r$-values around 1 can be reached. 
Also the data point for $r_0\sigma$  [\Eq{eq:r0sigma_meas}] then 
agrees with the $O(p^4)$
BChPT expression for $\sigma(m_\pi)$ [\Eq{eq:sigma_exp}], whether or 
not the data point for $r_0\sigma$ was included in the fit (compare
Figs.~\ref{fig:Splots} and \ref{fig:Nplot}). For the fit range 
$(r_0m_\pi)^2<3.0$ this is not the case anymore.

\subsection{Weighted averages}

Based on the results summarized in \Tab{tab:fitparaS}, we estimate the
weighted averages of our fit results for $r_0\sigma_{\mathrm{phys}}$
and $r_0M^{\mathrm{phys}}_N$. For the weights we use the statistical
error, and only values from fits with $\chi^2_r<1.3$ are allowed to enter the
average. We obtain the values listed in \Tab{tab:weighted_ave}. As one
can see from this table, our results for $r_0$ for $(r_0m_\pi)^2<1.6$
and $(r_0m_\pi)^2<1.3$ are consistent within errors, including that
from the stand-alone fit.

In this table we also give the systematic error due to varying $c_3$ (second
parenthesis) and $\bar{l}_3$ (third parenthesis) one standard deviation around
their phenomenological values in Eqs.~\eqref{eq:l3bar}
and \eqref{eq:c2c3_Meissner}. In total the systematic error is as large as the
statistical error.

\begin{table}[h]
\begin{center}\small
\begin{tabular}{l@{\;}l@{}c@{\quad}l@{\quad}l@{\quad}l@{\quad}l}
\hline\hline
 \multicolumn{2}{l}{Fit} & $(r_0m_\pi)^2_{\max}$ & $r_0\sigma_{\mathrm{phys}}$  &
$r_0$\,[GeV$^{-1}$] &
$r_0$\,[fm]
& 
$\sigma_{\mathrm{phys}}$ [MeV]\\
\hline\\*[-2ex]
 $O(m_\pi^4):$  & $\chi^2_N$
& $1.6$ &  0.103(23)(5) & 2.51(6)(2) & 0.495(12)(4) & 41(9)(2) \\
$O(m_\pi^4):$ & $\chi^2_{N\sigma}$ 
& $1.6$ & 0.095(11)(12)(4) & 2.54(3)(4)(2) & 0.501(6)(8)(4) & 37(4)(5)(2) \\
$O(m_\pi^4):$ & $\chi^2_{N\sigma}$  
& $1.3$ & 0.093(20)(15)(5) & 2.54(5)(5)(2) & 0.501(10)(10)(4) & 37(8)(6)(2) \\
$O(m_\pi^3):$ & $\chi^2_{N\sigma}$ 
& $1.0$ & 0.121(5) & 2.49(3) & 0.491(6) & 49(2) \\
 $O(m_\pi^2):$ & $\chi^2_{N\sigma}$ & $1.0$ & 0.065(7) & 2.58(3) &
0.509(5) &
25(3) \\
\hline\hline
\end{tabular} 
\end{center}
\caption{Weighted averages of fit results with $\chi^2_r<1.3$. The
first row gives averages for (stand-alone) fits to the nucleon mass
data; the remaining rows for (combined) fits to the nucleon mass and
$\sigma$-term data. The first column specifies the order of the chiral
expansion, the second the upper limit on $r_0m_\pi$. Some of the
weighted averages come with a statistical and systematic error: The
error in the first parenthesis is always the statistical error and the
second (third) the systematic error estimated by changing the fixed
parameter $c_3$ ($\bar{l}_3$) by one standard deviation [see
\Eq{eq:l3bar} and \eqref{eq:c2c3_Meissner}]. The numbers that went
into the averages are listed in Tables \ref{tab:fitparaS} and
\ref{tab:fitparaN}.}
\label{tab:weighted_ave}
\end{table}

\subsection{Fits to lower order expansions}
\label{sec:loworderfits}

It is interesting to check the robustness of the above estimates
for $r_0$ and $\sigma_{\mathrm{phys}}$ by fits to $O(m_\pi^2)$ and
$O(m_\pi^3)$ BChPT. Up to these orders, only $c_1$ and $M_0$ are left
as free parameters. Note that we still have to correct for the
finite-volume effect in the nucleon mass data. We do this by setting
(as above) $c_2=3.3\,\mathrm{GeV}^{-1}$ and $c_3=-4.7\,\mathrm{GeV}^{-1}$ 
in $\Delta M_N$.

For the fits to $O(m_\pi^2)$ and $O(m_\pi^3)$ BChPT we employ the
$\chi^2$-function for our combined fits [\Eq{eq:chi2Mnsigma}]. The
fitting ranges are chosen as above but we add to these
\mbox{$(r_0m_\pi)^2_{\max} = 1.0$}. It turns out that only for
$(r_0m_\pi)^2_{\max}=1.0$ reasonable fits to $O(m_\pi^2)$ and
$O(m_\pi^3)$ BChPT can be found.

\begin{figure}
 \centering
 \mbox{\includegraphics[width=0.45\linewidth]{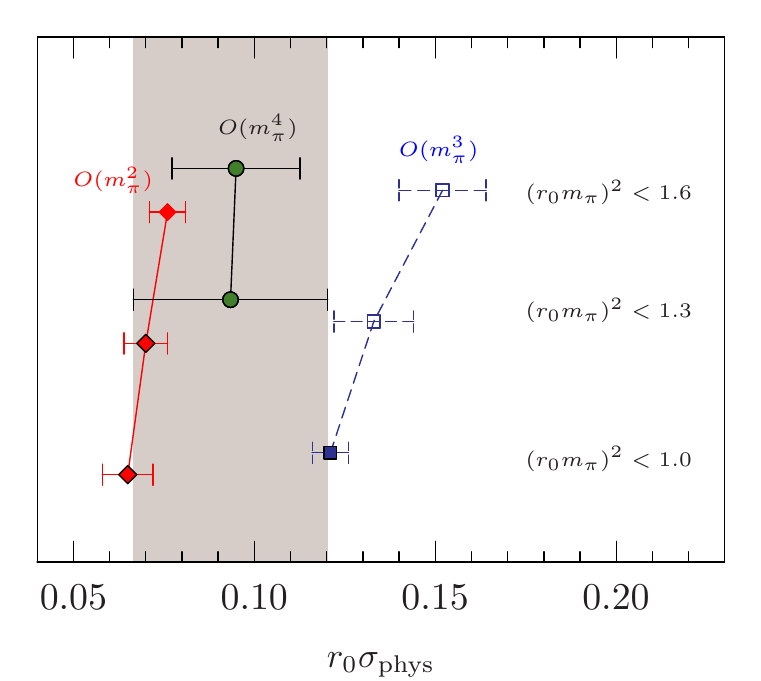}
 \includegraphics[width=0.455\linewidth]{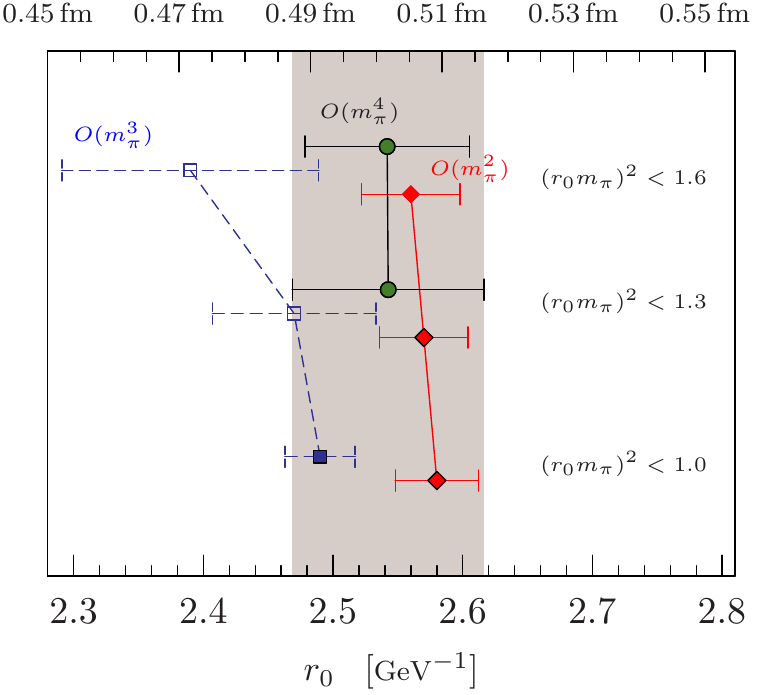}}
  \caption{Results for $r_0\sigma$ at the physical point (left panel)
  and $r_0$ (right panel) from fits to BChPT truncating the fit 
  functions $M_N(m_\pi)$ and $\sigma(m_\pi)$ at different orders
   in $m_\pi$. Only results for our combined fits [\Eq{eq:chi2Mnsigma}] are
   shown. Points are grouped together with respect to the upper
   cutoff on $(r_0m_\pi)^2$. Red diamonds are for fits to order $O(m_\pi^2)$, blue
   squares are for $O(m_\pi^3)$. Green circles refer to the weighted averages
   in \Tab{tab:weighted_ave}, where for the error the
    statistical and systematic errors have been added in quadrature. 
    Full (open) symbols refer to fits for which $\chi^2_r<2.0$ ($>2.0$); 
    black-framed full symbols to fits where $\chi^2_r<1.3$. 
    The gray bands represent the errors for the lower green circle, which
    corresponds to our final values for $r_0\sigma_{\mathrm{phys}}$
    (left) and $r_0$ (right).}
  \label{fig:r0sigma_r0_r0mpi2_max_diffO}
\end{figure}

Our results for $r_0\sigma_{\mathrm{phys}}$ and $r_0$ from these fits are listed in
\Tab{tab:weighted_ave}. To ease the comparison we also show them with
our $O(m_\pi^4)$ results in
\Fig{fig:r0sigma_r0_r0mpi2_max_diffO}. Open (full) symbols correspond
to fits where $\chi^2_r>2$ ($\chi^2_r<2$), black-framed full symbols
represent good fits where $\chi^2_r<1.3$.

As can been seen from this figure, fits to different orders in $m_\pi$
result in slightly different estimates both for $r_0$ and
$r_0\sigma_{\mathrm{phys}}$, 
and also come with a different fit quality. However, these deviations
get smaller when increasing the order or decreasing
$(r_0m_\pi)^2_{\max}$. For fixed $(r_0m_\pi)^2_{\max}$, points for
different orders seem to alternate around a yet unknown value with the
tendency of coming closer to that with each order. Results from fits
to even larger orders or lower $(r_0m_\pi)^2_{\max}$ will likely be
found within the error bounds of the $O(m_\pi^4)$ results.

We therefore conclude that our estimates for $r_0$ and
$\sigma_{\mathrm{phys}}$, that is, 
the weighted averages of our results from the combined $O(m_\pi^4)$ fits with
$(r_0m_\pi)_{\max}^2=1.3$, lead to sufficiently conservative errors
to accommodate all of the uncertainties involved when fitting nucleon
mass data to BChPT.

\subsection{Comparing different orders}

\begin{figure}[tb]
\centering
 \includegraphics[width=0.8\linewidth]{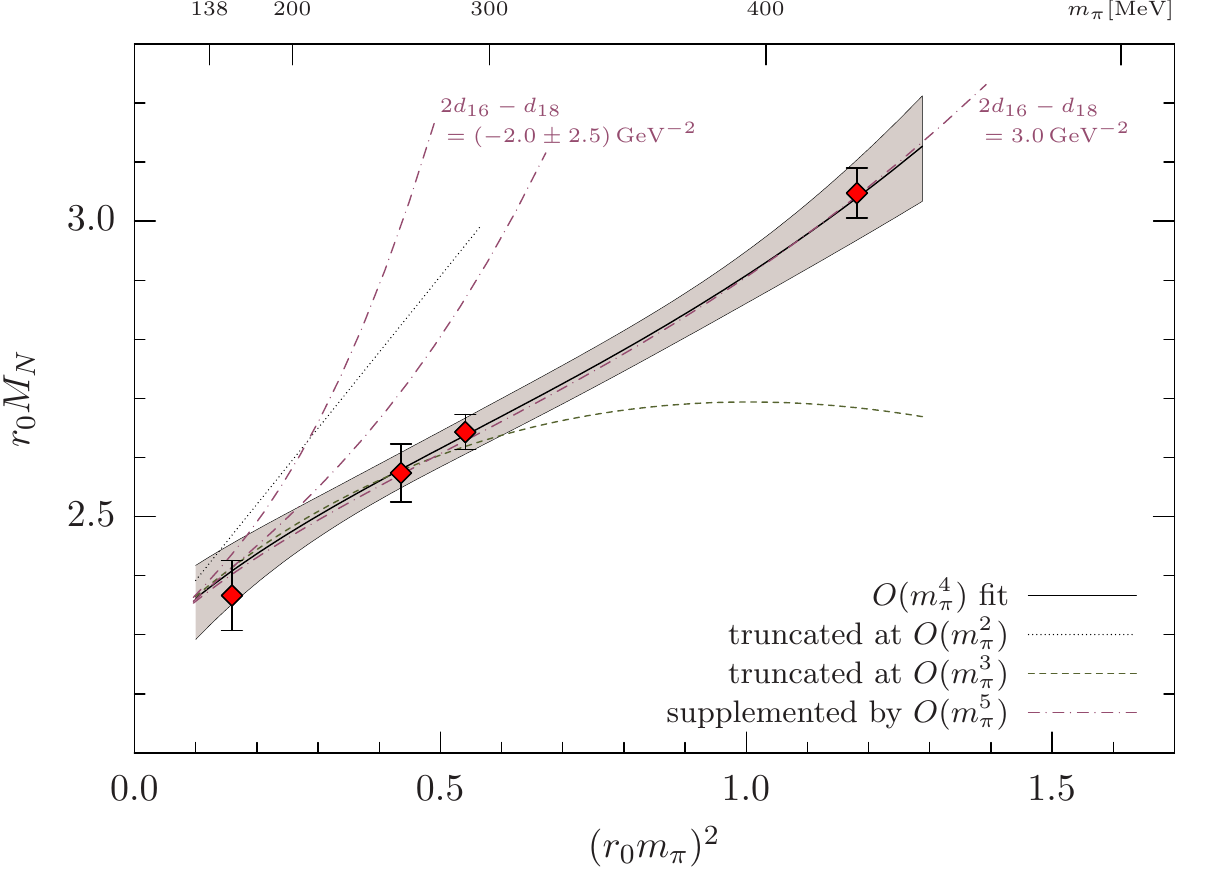}
 \caption{Comparison of one of our combined fits  to
         the contributions from lower orders ($O(m^2_\pi)$ and $O(m^3_\pi)$) and
         to the $O(m^5_\pi)$ expansion. For all expansions the parameters are
         fixed to those of the fit (\texttt{Sfo1} in
         \Tab{tab:fitparaS}). The additional parameters
         for the $O(m^5_\pi)$ expansion were fixed to values consistent
         with current expectations (see text). In addition,
         we show the $O(m^5_\pi)$ expansion assuming a larger value
         for $2d_{16}-d_{18}$ than currently expected from
         phenomenology (very right dashed-dotted line). 
         For simplicity, we show only finite-volume corrected data
         (diamonds) from the largest lattice volumes, even though for
         $(r_0m_\pi)^2=0.435$ and $0.540$ also points from smaller
         volumes entered the fit.}   
  \label{fig:ComparingDiffOrders}
\end{figure}

Let us finally try to get an impression of the convergence properties
of the BChPT formulae we are using. To this end we plot in
Fig.~\ref{fig:ComparingDiffOrders} a combined fit (\texttt{Sfo1} in
\Tab{tab:fitparaS}, where $c_3$ is a free parameter) along with
the curves which result from truncating the BChPT function at
$O(m^2_\pi)$ and at $O(m^3_\pi)$, using the same parameter values. In
addition we show two (three) curves where the contribution of fifth order in
$m_\pi$ has been added to the fitted function varying the new LECs
$d^r_{16}$, $d^r_{18}$ and $l^r_4$ appearing in this 
contribution within a phenomenologically acceptable (slightly expanded)
range. At this order, the LECs $d^r_{16}$ and $d^r_{18}$
enter as the difference $2d^r_{16}-d^r_{18}$ (see \ref{app:BChPT} for
details), which currently is known only approximate,
$2d^r_{16}-d^r_{18}=(-2.0\pm2.5)\mathrm{GeV}^{-2}$
\cite{Fettes:2000fd,Bernard:2006te}. For $l^r_4$ we use the value for
$\bar{l}_4$ given in \cite{Colangelo:2010et}. 

Given this range of expected values, we see that the fifth order 
correction becomes a non-negligible effect already at pion masses well
below the physical kaon mass. More specifically, at pion masses of
$\sim 350\,\mathrm{MeV}$, the fifth order contribution is already of
about the same size as the third order term from the leading-one-loop
correction. This leads us to the conclusion that, in case of the
nucleon mass and sigma term, BChPT (with the current LECs) shows no
sign of convergence beyond $m_\pi > 250\,\mathrm{MeV}$.
Discussions pointing in this direction can also be found in, e.g., 
\cite{Leinweber:1999ig,Beane:2004ks,Leinweber:2005xz,McGovern:2006fm,Schindler:2007qe}.
If one allowed, however, for a slightly wider range for
$2d^r_{16}-d^r_{18}$, say $2d^r_{16}-d^r_{18}=3.0\,\mathrm{GeV}^{-2}$,
the fifth order contribution could be much smaller (see
Fig.~\ref{fig:ComparingDiffOrders}). But this is speculative
only, and an improved knowledge on the values for $d^r_{16}$ and $d^r_{18}$ is
required to decide on that. At least from the small difference of the
$O(p^3)$ and $O(p^4)$ functions at small $(r_0m_\pi)^2$, we learn that
the LECs $c_2$, $c_3$ and $e_1^r$ are not well constrained by nucleon
mass data at pion masses $m_\pi<300$\,MeV.

\section{Conclusions}
\label{sec:conclusion}

We have presented $N_f=2$ QCD nucleon mass data.  The corresponding
pion mass values range from about $1.5\,\textrm{GeV}$ down to
$157$~MeV. To estimate the nucleon $\sigma$-term we have performed two
kinds of fits to expressions from $O(p^4)$ BChPT: (stand-alone) fits
to our nucleon mass data and simultaneous fits to the nucleon mass and
$\sigma$-term data. The latter was determined in a separate study
\cite{Bali:2011ks} at a pion mass of about $290\,\textrm{MeV}$.

For the fits, different fitting ranges in $m_\pi$ and $L$ (spatial
lattice extension) have been tested. We find that if one demands
$m_\pi < 500\,\textrm{MeV}$ [$(r_0m_\pi)^2<1.6$] and also
$L>1.5\,\textrm{fm}$ acceptable fits to $O(p^4)$
BChPT can be found.\footnote{Still one needs to have access to
  $m_\pi$-values for which $m_\pi L\ge3.5$. Otherwise the finite-volume 
  effect for $m_\pi$ is not under control.} 
These fits do not only give a good description of 
the $m_\pi$ dependence of the data but also of the finite-volume
effects. Generally, our simultaneous fits perform better than fits to
the nucleon mass data alone. They are also robust against variations of
$(r_0m_\pi)^2_{\max}$ and of the low energy constants $c_2$, $c_3$ and
$\bar{l}_3$. We have found a strong correlation between the counterterm
coefficient $e^r_1$ and $c_3$. More data points below 500\,MeV pion
mass will be needed to resolve this issue or to fix $c_2$ and $c_3$
through lattice data.

\begin{figure}[t]
  \centering
  \includegraphics[width=0.9\textwidth]{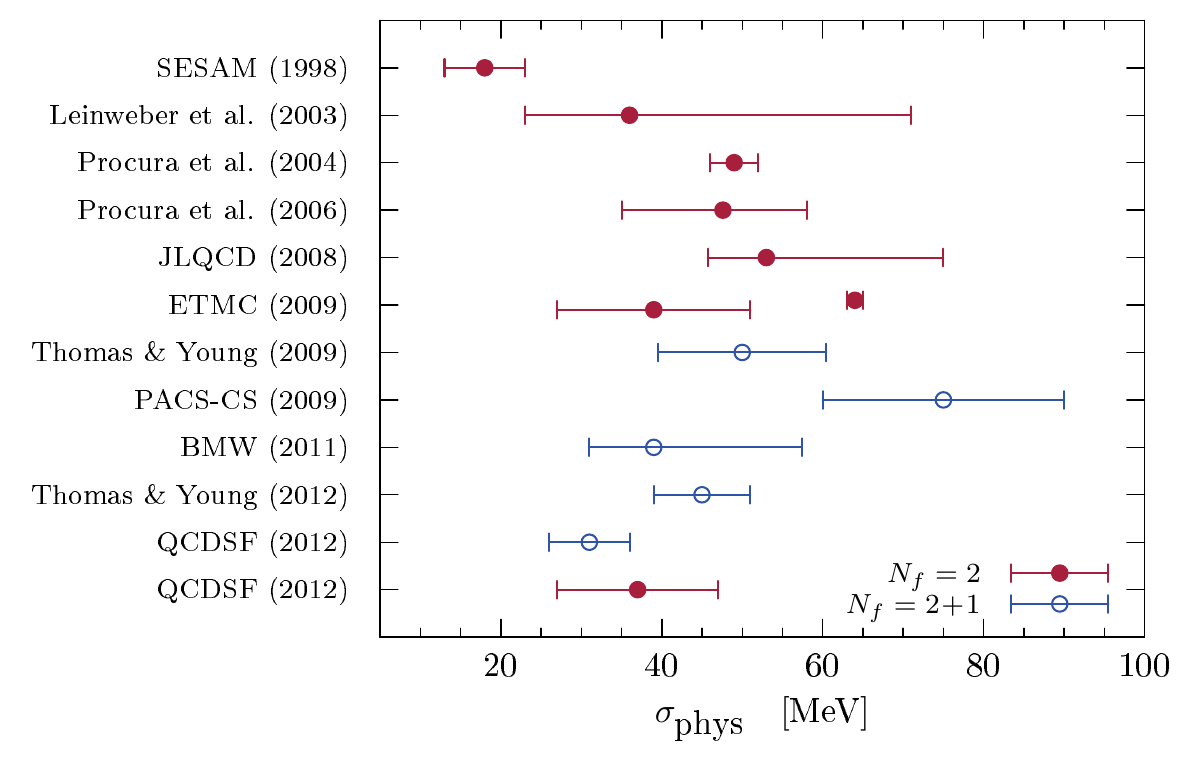}
  \caption{Comparison of lattice estimates for the pion-nucleon
    $\sigma$-term at the physical point for $N_f=2$ and $N_f=2+1$
    QCD
    \cite{Gusken:1998wy,Procura:2003ig,Procura:2006bj,Ohki:2008ff,Young:2009zb,
      Ishikawa:2009vc,Shanahan:2012wh,Horsley:2011wr}. The lowermost
    point represents our estimate for $\sigma_{\mathrm{phys}}$ 
   [\Eq{eq:estimate_sigma_13_final}].}
  \label{fig:sigma_lattice_results}
\end{figure}

As our final estimates we quote
\begin{equation}
\label{eq:estimate_r0_13_final}
 r_0 = 0.501(10)(11)\,\text{fm}
\end{equation}
and
\begin{equation}
\label{eq:estimate_sigma_13_final}
 \sigma_{\mathrm{phys}}  = 37(8)(6)\,\textrm{MeV}\,.
\end{equation}
These numbers are the weighted averages given in the last row of
\Tab{tab:weighted_ave}, adding the two systematic errors in
quadrature. These numbers result from our fits to $O(p^4)$ BChPT,
fitting simultaneously the nucleon mass and $\sigma$-term data up to
pion masses of about $433\,\textrm{MeV}$. Within errors, we find these
numbers to be consistent with corresponding fits to $O(p^2)$ and
$O(p^3)$ BChPT (see \Fig{fig:r0sigma_r0_r0mpi2_max_diffO}), and also
with our recent estimate $\sigma_{\mathrm{phys}}=31(3)(4)\,\text{MeV}$
for $N_f=2+1$ \cite{Horsley:2011wr}.

\begin{figure*}
 \centering
   \includegraphics[width=0.8\linewidth]{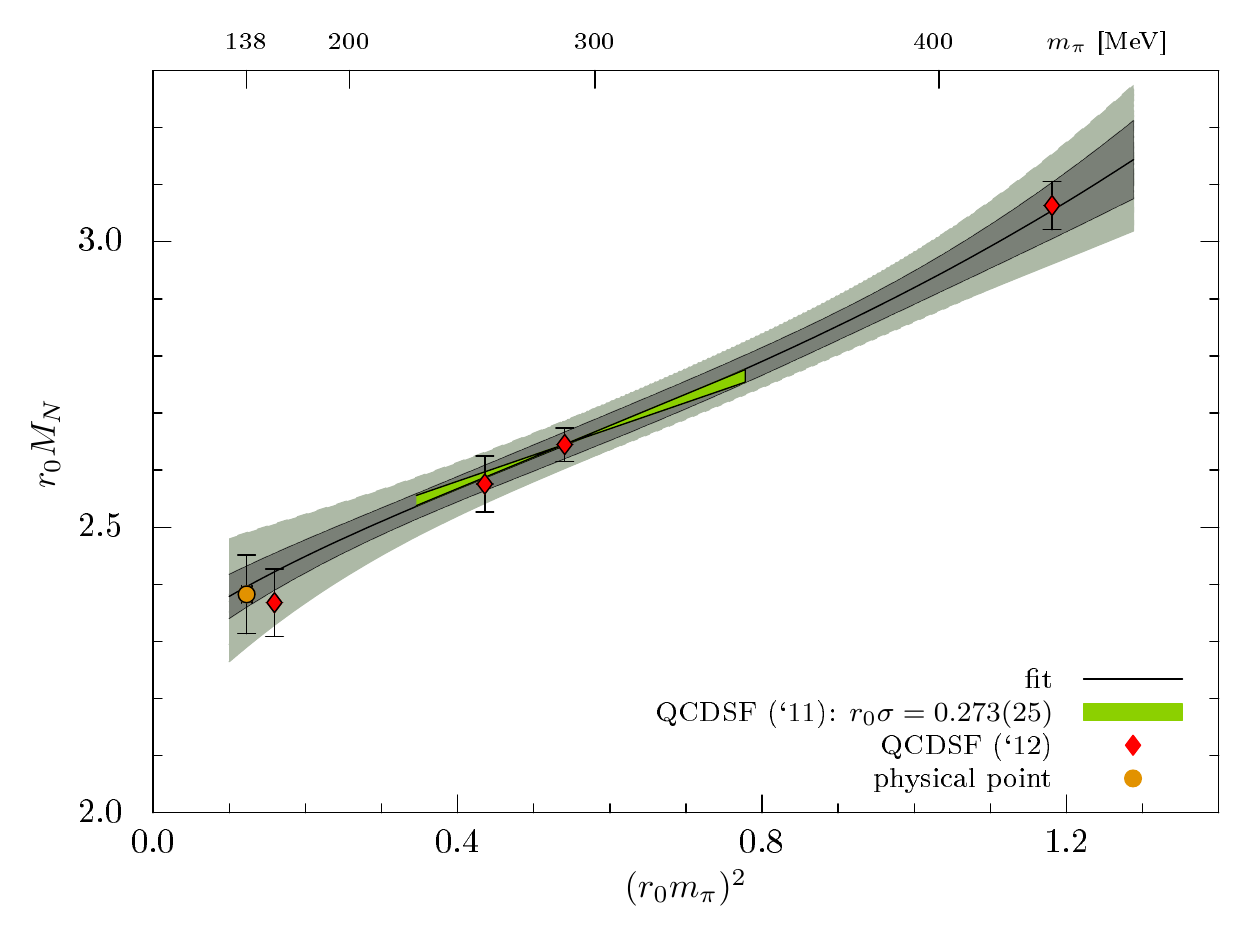}
   \includegraphics[width=0.8\linewidth]{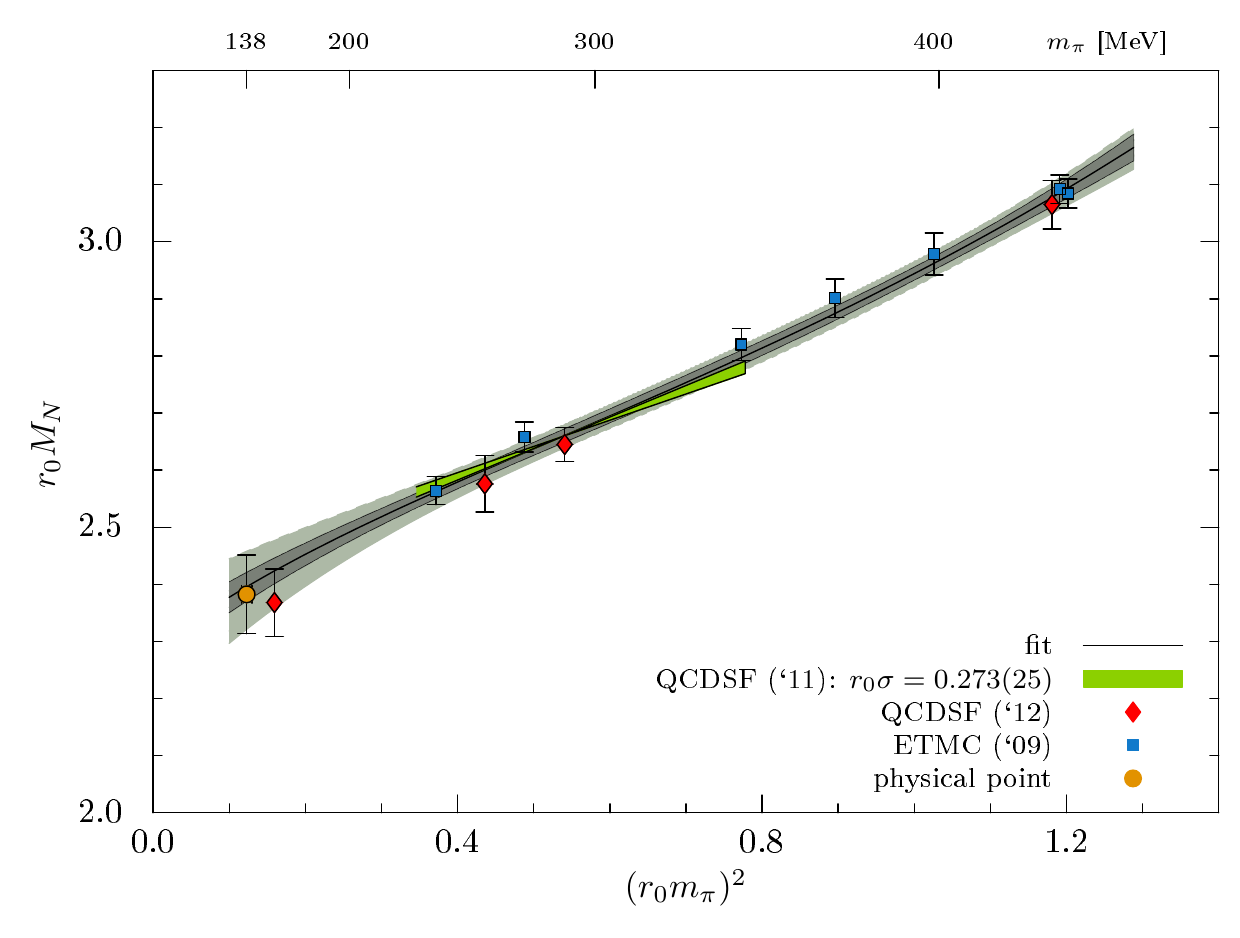}
  \caption{Top: combined fit to our nucleon mass (red
         diamonds) and $\sigma$-term data (light green band). Shown
         are the volume-corrected 
         data (determined through the fit) for one volume only. The
         fit's characteristics (fit 
         ranges, fixed parameters) are the same as for the fit labeled
         \texttt{Soo1}
         in \Tab{tab:fitparaS}. The inner band is the error band for
         the fit, the outer band illustrates the variation of the
         fit changing $\bar{l}_3$ and $c_3$ by one standard deviations around
         their phenomenological values [Eqs.~\eqref{eq:l3bar} and
         \eqref{eq:c2c3_Meissner}]. For the physical point, marked by a
         (yellow) circle, we use our final estimate for $r_0$
         [\Eq{eq:estimate_r0_13_final}]. Bottom: same as top panel, but the
         combined fit also includes the nucleon mass data of
         the ETM Collaboration (blue squares)
         \cite{Alexandrou:2009qu}. Note that their data is plotted
         against the $\pi^+$ mass.}
  \label{fig:QCDSF_vs_ETMC}
\end{figure*} 

Agreement is also found if one compares with other estimates,
for example, with the $N_f=2+1$ estimate of
the BMW Collaboration \cite{Durr:2011mp}, the CSSM
results for $N_f=2$ and $N_f=2+1$
\cite{Leinweber:2003dg,Young:2009zb,Shanahan:2012wh} or with one of the ETM
estimates for $N_f=2$ \cite{Alexandrou:2009qu} (see
\Fig{fig:sigma_lattice_results} for a comparison). Note that ETM
quotes a final result $\sigma_{\mathrm{phys}}=64(1)\textrm{MeV}$ from a fit to
$O(p^3)$ BChPT, however they 
find also $\sigma_{\mathrm{phys}}=39(12)\textrm{MeV}$ from a modified fit function
(see Figs.~13 and 14 of \cite{Alexandrou:2009qu}). Also from our fits
to $O(p^3)$ BChPT we could find larger values for $\sigma$, depending
on the largest pion mass included in the fits. However, based on the
systematic procedure we have applied here, we think the value in
\Eq{eq:estimate_sigma_13_final} is more reliable.

To compare our results with other recent $N_f=2$ calculations we
have redone our combined fits including the raw (i.e., not corrected
for finite-size effects) data of the ETM Collaboration
\cite{Alexandrou:2009qu} into our analysis. This roughly doubles
the amount of data which is available at $m_\pi<300$\,MeV.
Repeating fits for different fixed parameter values $\bar{l}_3$, $c_2$ and
$c_3$ we obtain values for $r_0$ and $\sigma_{\mathrm{phys}}$ which agree within
errors with those of Eqs.~\eqref{eq:estimate_r0_13_final} and
\eqref{eq:estimate_sigma_13_final}. The bottom panel of
\Fig{fig:QCDSF_vs_ETMC} shows one of these fits:
Red diamonds represent our data, blue squares ETMC's 
points, all after subtracting the finite-volume corrections. As above,
these corrections were determined directly through the
fit, using points from different volumes but same $(r_0m_\pi)^2$. For
simplicity though, in this figure only points from the largest
available lattice volumes are shown. The short (green) band is from the  
value $r_0\sigma=0.273(25)$ \cite{Bali:2011ks} which constrains the
slope of the fitting function at $r_0m_\pi=0.735$. The yellow circle
marks the physical point using our $r_0$-value in
\Eq{eq:estimate_r0_13_final}. For a comparison, the top
panel of \Fig{fig:QCDSF_vs_ETMC} shows the corresponding fit
(labeled \texttt{Soo1} in \Tab{tab:fitparaS}) without ETMC's
points included.

We conclude with a note on $r_0$: The value $r_0 = 0.501(10)(11)\,\text{fm}$
obtained from our fits is somewhat surprising.
In other studies (including ours) $r_0$ was often found to be about
0.47\,fm (see, e.g., \cite{Aubin:2004wf,Gockeler:2005mh,Gockeler:2005rv,Alexandrou:2009qu}). Here $r_0$ was fixed iteratively forcing the fit to be
self-consistent, i.e. $r_0=\widehat{M}_N(r_0\cdot
m_{\mathrm{phys}})/M^{\mathrm{phys}}_N$. Our value for $r_0$ (and that for
$\sigma_{\mathrm{phys}}$) is thus valid as long as the following assumptions are
satisfied: (a) $O(m_\pi^4)$ BChPT is sufficient to describe our data below
$m_\pi=435$ MeV, (b) discretization effects are really negligible and (c) a
physical value of 938 MeV is adequate for a two-flavor calculation. Whether
these assumptions have to be changed or relaxed has to be seen 
when there will be more data available at smaller pion masses and 
lattice spacings, and also for simulations with
$N_f=2+1$ and $N_f=2+1+1$. From the right panel of 
\Fig{fig:r0sigma_r0_r0mpi2_max_diffO}, however, we see a trend for
$r_0$: it increases when the upper limit on $r_0m_\pi$ is lowered. 

Note that also a recent study \cite{Fritzsch:2012wq} of the kaon decay constant
finds a $r_0$-value of around 0.5\,\text{fm} from a two-flavor lattice
calculation. Their estimate is completely consistent with our estimate of
\Eq{eq:estimate_r0_13_final}.

\section*{Acknowledgements}

This work was supported by the European Union under the Grant
Agreement numbers 238353 (ITN STRONGnet), 256594 (IRG), 283286 (Hadron
Physics 3), and by the Deutsche Forschungsgemeinschaft SFB/Transregio
55. S.~Collins acknowledges support from the
Claussen-Simon-Foundation (Stifterverband f\"ur die Deutsche
Wissenschaft). J.~Zanotti was supported by the Australian Research 
Council under grant FT100100005. B~Gl\"a\ss{}le and N.~Najjar received
support from the EU Research Infrastructure Action HPC-Europa2 228398. 
Computations were performed on the
SFB/TR55 QPACE supercomputers, the BlueGene/P (JuGene) and the Nehalem
Cluster (JuRoPA) of the J\"ulich Supercomputer Center, the IBM
BlueGene/L at the EPCC (Edinburgh) and the SGI Altix ICE machines at
HLRN (Berlin/Hannover). We thank the support staffs of these
institutions. The Chroma software suite \cite{Edwards:2004sx} and
BAGEL \cite{Bagel} was used in this work and gauge configurations were
generated using the BQCD code \cite{Nakamura:2010qh} on QPACE and
BlueGenes.

\appendix

\section{Nucleon mass and pion-nucleon $\sigma$-term from BChPT}
\label{app:BChPT}

In this appendix we derive our fit formulae from a next-to-leading 
one-loop order (or ${\cal O}(p^4)$) calculation of the nucleon
mass in covariant baryon chiral perturbation theory (BChPT) for two 
light flavors. The pion-nucleon $\sigma$ term then follows from an 
application of the Feynman-Hellmann theorem (see Eq.(2)). In addition 
we need to know the connection between the quark mass and the pion mass to 
${\cal O}(p^4)$~\cite{Gasser:1983yg,Bellucci:1994eb}. Denoting the 
mass of the degenerate light quarks by  $m_\ell = m_u = m_d$ the leading 
term is given by the Gell-Mann--Oakes--Renner relation
\begin{equation}
m_\pi^2 = 2 B_0 m_\ell \,.
\end{equation}
While the quark mass $m_\ell$ and the parameter $B_0$ are scheme and 
scale dependent, the auxiliary variable 
\begin{equation}
{\overline m}^2 \equiv 2 B_0 m_\ell   \label{eq:GOR}
\end{equation}
is independent of these conventions. 

The generic ${\cal O}(p^4)$ result for the nucleon mass can be written as 
\begin{equation}
M_N=M_0+M^{(1)}+M^{(2)}+M^{(3)}+M^{(4)}+{\cal O}(p^5) \,. \label{eq:mfull}
\end{equation}
Evaluating the required loop diagrams with 
$\overline{\mathrm{IR}}$-regularization \cite{Dorati:2007bk} one obtains
\begin{align}
M^{(1)}=& \, 0 \,, \label{eq:m1} \\
M^{(2)}=& -4 c_1 {\overline m}^2 \,, \label{eq:m2}\\
M^{(3)}=& -\frac{3 (g_A^0)^2\,{\overline m}^3}{16\pi^2(F_\pi^0)^2}
   \left\{ \sqrt{1-\frac{{\overline m}^2}{4M_0^2}}
                      \arccos{\frac{{\overline m}}{2M_0}}
    +\frac{{\overline m}}{4M_0}
         \log\frac{{\overline m}^2}{M_0^2}\right\} \,, \label{eq:m3} \\
M^{(4)}=& \, 4 \tilde{e}_1^{r}(\lambda)\,{\overline m}^4
  +\frac{3 {\overline m}^4}{64\pi^2(F_\pi^0)^2}
       \log\frac{{\overline m}^2}{\lambda^2}
   \left\{8c_1-c_2-4c_3-\frac{(g_A^0)^2}{M_0}\right\} \nonumber \\
       & {} +\frac{3 {\overline m}^4}{64\pi^2(F_\pi^0)^2}
  \left\{\frac{c_2}{2}-3 \frac{(g_A^0)^2}{M_0}
   +\frac{(g_A^0)^2}{M_0} \log\frac{{\overline m}^2}{M_0^2} \right\}\nonumber \\
       & {}-\frac{3 c_1 (g_A^0)^2\,{\overline m}^6}{16\pi^2(F_\pi^0)^2M_0^2} 
         \left\{ \log\frac{{\overline m}^2}{M_0^2}
         -\frac{{\overline m}}{M_0}\,\frac{\arccos{\frac{{\overline m}}{2M_0}}}
         {\sqrt{1-\frac{{\overline m}^2}{4M_0^2}}} \right\} \,. \label{eq:m4}
\end{align}
Here $g_A^0$ and $F_\pi^0$ denote the axial coupling constant of the
nucleon and the pion decay constant in the chiral limit, and $c_1$,
$c_2$, $c_3$ are the standard low-energy constants (see, e.g.,
\cite{Becher:2001hv}). The renormalized counterterm coefficient
$\tilde{e}_1^{r}(\lambda)$ depends on the scale $\lambda$ of
dimensional regularization in such a way that $M^{(4)}$ does not
depend on $\lambda$.

The generic next-to-leading one-loop result for the sigma term of the 
nucleon reads
\begin{equation}
\sigma = \sigma^{(1)}+\sigma^{(2)}+\sigma^{(3)}+\sigma^{(4)}
               +{\cal O}(p^5) \,. \label{eq:totalsigma}
\end{equation}
Utilizing the relation
\begin{equation}
\sigma = m_\ell \frac{\partial M_N (m_\ell)}{\partial m_\ell} 
       = {\overline m}^2 \,\frac{\partial M_N({\overline m}^2)}
                                {\partial {\overline m}^2}  \label{eq:FH}
\end{equation}
one obtains from Eqs.(\ref{eq:m1})-(\ref{eq:m4})
\begin{align}
\sigma^{(1)} =& \, 0 \,, \label{eq:s1} \\
\sigma^{(2)} =& -4 c_1 {\overline m}^2 \,, \label{eq:s2}\\
\sigma^{(3)} =& -\frac{3 (g_A^0)^2\,{\overline m}^3}{16\pi^2(F_\pi^0)^2}
 \left\{ \frac{3-\frac{{\overline m}^2}{M_0^2}}
          {2\sqrt{1-\frac{{\overline m}^2}{4M_0^2}}}
          \arccos{\frac{{\overline m}}{2M_0}}
       +\frac{{\overline m}}{2M_0}
          \log\frac{{\overline m}^2}{M_0^2}\right\} \,, \label{eq:s3}\\
\sigma^{(4)} =& \, 8 \tilde{e}_1^{r}(\lambda)\,{\overline m}^4
     +\frac{3\,{\overline m}^4}{32\pi^2(F_\pi^0)^2}
       \log\frac{{\overline m}^2}{\lambda^2}
       \left\{8 c_1-c_2-4 c_3-\frac{(g_A^0)^2}{M_0}\right\} \nonumber \\
       & {}+\frac{3 {\overline m}^4}{32\pi^2(F_\pi^0)^2}
          \left\{4 c_1-2 c_3-3 \frac{(g_A^0)^2}{M_0}
         +\frac{(g_A^0)^2}{M_0} \log\frac{{\overline m}^2}{M_0^2} \right\}
                                                            \nonumber \\
       & {} -\frac{3 c_1\,(g_A^0)^2\,{\overline m}^6}{16\pi^2(F_\pi^0)^2M_0^2}
          \left\{\frac{1}{1-\frac{{\overline m}^2}{4M_0^2}}
           +3\log\frac{{\overline m}^2}{M_0^2}
           -\left(\frac{7}{2}-\frac{3 {\overline m}^2}{4M_0^2}\right)
          \frac{{\overline m}}{M_0}\,
           \frac{\arccos{\frac{{\overline m}}{2M_0}}}
                {\left( 1-\frac{{\overline m}^2}{4M_0^2} \right)^{3/2}} 
                                                    \right\} \,. \label{eq:s4}
\end{align}

Finite volume corrections to the nucleon mass can be evaluated from
the same Feynman diagrams. At next-to-leading one-loop order one 
gets~\cite{AliKhan:2003cu}
\begin{equation}
\Delta M_N(\overline{m}^2,L) = \Delta M^{(3)}(\overline{m}^2,L)
   + \Delta M^{(4)}(\overline{m}^2,L)+{\cal O}(p^5)
\end{equation}
with 
\begin{align}
\Delta M^{(3)} &=  \frac{3 (g_A^0)^2 M_0 \overline{m}^2}{16 \pi^2 (F_\pi^0)^2}
 \int_0^\infty \! \mathrm d x \,
 \sum_{\vec{n} \neq \vec{0}} \,
  K_0 \left( L |\vec{n}| \sqrt{M_0^2 x^2 + \overline{m}^2 (1-x)} \right) 
    \,, \label{eq:deltam3}\\
\Delta M^{(4)}&=  \frac{3 \overline{m}^4}{4 \pi^2 (F_\pi^0)^2} 
  \sum_{\vec{n}\neq \vec{0}} \left[ (2 c_1 - c_3) 
       \frac{K_1(L |\vec{n}| \overline{m})}{L |\vec{n}| \overline{m}}
     + c_2 \frac{K_2(L |\vec{n}| \overline{m})}
               {(L |\vec{n}| \overline{m})^2} \right] \,,\label{eq:deltam4}
\end{align}
where $K_i$ is a modified Bessel function. Note that the value 
$\vec{n} = \vec{0}$ is omitted in the threefold sum over the integers
$n_1$, $n_2$, $n_3$.

For the applications in the present paper it is advantageous to consider 
the nucleon mass as a function of the pion mass $m_\pi$ (the mass of the 
lowest lying $0^-$ state in the simulation). Therefore we have to convert
our expressions for $M_N({\overline m}^2)$ of Eq.(\ref{eq:mfull}) into 
expressions for $M_N(m_\pi^2)$. Utilizing the ${\cal O}(p^4)$ 
result~\cite{Gasser:1983yg}
\begin{equation}\label{eq:mpi}
m_\pi^2 = {\overline m}^2 
  + 2 l_3^{r}(\lambda)\,\frac{{\overline m}^4}{(F_\pi^0)^2}
  + \frac{{\overline m}^4}{32\pi^2(F_\pi^0)^2}
          \log\frac{{\overline m}^2}{\lambda^2} +{\cal O}(p^6)
\end{equation}
we can eliminate the dependence on ${\overline m}$ at the cost of 
introducing the renormalized low-energy constant $l_3^{r}(\lambda)$, 
which cancels the dependence on the renormalization scale $\lambda$ 
of the associated chiral logarithm in Eq.~\eqref{eq:mpi}. 
Note, however, that $l_3^{r}(\lambda)$ can be subsumed into an 
effective coupling $e_1^{r}(\lambda)$ via
\begin{equation}\label{eq:e1mod}
e_1^{r}(\lambda)\equiv\tilde{e}_1^{r}(\lambda)+2\,l_3^{r}(\lambda)\,
\frac{c_1}{(F_\pi^0)^2}
\end{equation}
and therefore cannot be determined independently within an analysis 
of the mass of the nucleon. One finds
\begin{align}
M_N(m_\pi) =& \, M_0 -4 c_1\,m_\pi^2 \nonumber \\
   &{} -\frac{3 g_A^2\,m_\pi^3}{16\pi^2 F_\pi^2}
        \left\{\sqrt{1-\frac{m_\pi^2}{4M_0^2}}\arccos{\frac{m_\pi}{2M_0}}
         +\frac{m_\pi}{4M_0} \log\frac{m_\pi^2}{M_0^2}\right\} \nonumber \\
   & {} +4 e_1^{r}(\lambda)\,m_\pi^4
        +\frac{3 m_\pi^4}{64\pi^2 F_\pi^2}\log\frac{m_\pi^2}{\lambda^2}
         \left\{\frac{32}{3}\,c_1 - c_2-4 c_3-\frac{g_A^2}{M_0}\right\} 
                                                               \nonumber \\
   & {} +\frac{3 m_\pi^4}{64\pi^2 F_\pi^2}
         \left\{\frac{c_2}{2}- 3\, \frac{g_A^2}{M_0}
            +\frac{g_A^2}{M_0} \log\frac{m_\pi^2}{M_0^2} \right\}\nonumber \\
   & {} -\frac{3 c_1\,g_A^2\,m_\pi^6}{16\pi^2 F_\pi^2 M_0^2} 
         \left\{ \log\frac{m_\pi^2}{M_0^2}-\frac{m_\pi}{M_0}\,
          \frac{\arccos{ \frac{m_\pi}{2M_0}}}
               {\sqrt{1-\frac{m_\pi^2}{4M_0^2}}} \right\}
        +\delta M_N^{(5)} \,. \label{eq:mfinal}
\end{align}
Note that we have shifted the couplings 
$g_A^0\rightarrow g_A,\,F_\pi^0\rightarrow F_\pi$ to their physical 
values, consistent to the order at which we are working.
This applies also to the finite volume corrections, where we may in addition
replace ${\overline m}$ by $m_\pi$.

For the $\sigma$ term as a function of $m_\pi$ we obtain 
\begin{align}
\sigma (m_\pi) =& -4 c_1 m_\pi^2-\frac{3 g_A^2\,m_\pi^3}
               {16\pi^2 F_\pi^2}\left\{
        \frac{3-\frac{m_\pi^2}{M_0^2}}{2\sqrt{1-\frac{m_\pi^2}{4M_0^2}}}
        \arccos{\frac{m_\pi}{2M_0}}+\frac{m_\pi}{2M_0}\log\frac{m_\pi^2}{M_0^2}
               \right\} \nonumber\\
        &  {} +8 e_1^{r}(\lambda)\,m_\pi^4-\frac{8\,c_1l_3^{r}(\lambda)}
           {F_\pi^2}\,m_\pi^4+\frac{3\,m_\pi^4}{32\pi^2 F_\pi^2}\log
           \frac{m_\pi^2}{\lambda^2}\left\{\frac{28}{3}\,c_1-c_2-4\,c_3-
           \frac{g_A^2}{M_0}\right\} \nonumber \\
       & {} + \frac{3\,m_\pi^4}{32\pi^2 F_\pi^2}\left\{4\,c_1-2\,c_3-3\,
          \frac{g_A^2}{M_0}+\frac{g_A^2}{M_0}
          \log\frac{m_\pi^2}{M_0^2} \right\}\nonumber \\
       & {} -\frac{3\,c_1\,g_A^2\,m_\pi^6}{16\pi^2 F_\pi^2M_0^2}
          \left\{\frac{1}{1-\frac{m_\pi^2}{4M_0^2}}+3
          \log\frac{m_\pi^2}{M_0^2}-\left(\frac{7}{2}-\frac{3\,m_\pi^2}{4M_0^2}
          \right)\frac{m_\pi}{M_0}\,\frac{\arccos{\frac{m_\pi}{2M_0}}}
          {\left( 1-\frac{m_\pi^2}{4M_0^2} \right)^{3/2}} \right\} \nonumber \\
       & {} +\frac{5}{2}\,\delta M_N^{(5)} \,. \label{eq:sigmafinal}
\end{align}
Note that in contrast to the chiral extrapolation function of the mass 
of a nucleon given in \Eq{eq:mfinal} the dependence on the 
counter term $l_3^{r}(\lambda)$ introduced in \Eq{eq:mpi} cannot 
be absorbed completely into the effective coupling $e_1^{r}(\lambda)$. 
In order to calculate $\sigma (m_\pi^{phys})$ we therefore also 
need information on the numerical size of the coupling 
$\bar{l}_3$, which is related to the 
scale-dependent coupling $l_3^{r}(\lambda)$ via
\begin{equation}
l_3^{r}(\lambda) = - \frac{1}{64 \pi^2}
  \left( \bar{l}_3 + \log \frac{{\overline m}^2}{\lambda^2} \right) \,.
\end{equation}

Finally, expanding the expressions for $M_N (m_\pi)$ and 
$\sigma (m_\pi)$ given in Eqs.~\eqref{eq:mfinal} and \eqref{eq:sigmafinal}, 
respectively, in powers of $m_\pi$ up to ${\cal O}(m_\pi^4)$ we
arrive at the formulae Eqs.~\eqref{eq:Mn_exp} and \eqref{eq:sigma_exp} 
used in our fits.

Let us now estimate the theoretical uncertainty associated with our 
next-to-leading one-loop BChPT calculation of $M_N$ due to the 
truncation of ${\cal O}(p^5)$ effects (and all higher orders). 
A complete two-loop calculation of the nucleon mass in BChPT, 
employing a reformulated version of infrared regularization, 
has been worked out in Refs.~\cite{Schindler:2006ha, Schindler:2007qe}. 
Truncating the result at $\mathcal{O}(m_{\pi}^{5})$, we find
\begin{equation}\begin{split}
M_{N} =
M_{0}+\tilde{k}_{1}m_{\pi}^{2}+\tilde{k}_{2}m_{\pi}^{3}+\tilde{k}_{3}&m_{\pi}^{4
} \log\left(\frac{m_{\pi}}{\lambda}\right)+\tilde{k}_{4}m_{\pi}^{4}\\
&+\tilde{k}_{5
} m_{\pi}^{5}\log\left(\frac{m_{\pi}}{\lambda}\right)+\tilde{k}_{6}m_{\pi}^{5}
+\mathcal{O}(m_{\pi}^{6})\,.\end{split}
\end{equation}
The coefficients of this expansion are given by
\begin{eqnarray*}
\tilde{k}_{1} &=& -4c_{1}\,,\qquad 
\tilde{k}_{2} = -\frac{3(g^0_A)^{2}}{32\pi (F^0_\pi)^{2}}\,,\\
\tilde{k}_{3} &=& -\frac{3(g^0_A)^2 - 32c_{1}M_{0} + 3c_{2}M_{0} + 12c_{3}M_{0}}{32 M_{0}\pi^{2}(F^0_\pi)^2}\,,\\
\tilde{k}_{4} &=& 4e_{1}^{r}-\frac{3(2(g^0_A)^{2}-c_{2}M_{0})}
  {128 M_{0}\pi^{2}(F^0_\pi)^{2}}\,,\qquad 
\tilde{k}_{5} = \frac{3(g^0_A)^{4}}{64\pi^{3}(F^0_\pi)^{4}}\,,\\
\tilde{k}_{6} &=& \frac{3g^0_A}{256M_{0}^{2}\pi^{3}(F^0_\pi)^{4}}
\left((g^0_A)^{3}M_{0}^{2}+\pi^{2}\left(16 g^0_A M_{0}^{2}l_{4}^{r}
  +(F^0_\pi)^{2}(g^0_A-32M_{0}^{2}(2d_{16}^{r}-d_{18}^{r}))\right)\right)\,.
\end{eqnarray*}

The one-loop expression (coefficients $\tilde{k}_1$ -- $\tilde{k}_4$) 
is consistent with Eq.(\ref{eq:Mn_exp}).
In order to study the higher order effects we need values of
the additional low-energy constants
$l_{4}^{r}$ and $2d_{16}^{r}-d_{18}^{r}$. From
\cite{Fettes:2000fd,Bernard:2006te,Colangelo:2010et} we find 
$l_{4}^{r}(\lambda = 0.138\,\mathrm{GeV}) = 0.027$
and (with a considerable uncertainty) 
$d_{16}^{r}(\lambda = 0.138\,\mathrm{GeV})= -1.76\,\mathrm{GeV}^{-2}$.
For the scale-independent constant $d_{18}^{r}$, Ref.~\cite{Becher:2001hv}
derives the value $d_{18}^{r}=-0.80\,\mathrm{GeV}^{-2}$
from the Goldberger-Treiman relation. In view of the large uncertainty
of $d_{16}^{r}$ we use the rough estimate 
\mbox{$2 d_{16}^{r} -
  d_{18}^{r}=(-2.5\pm2.0)\,\textrm{GeV}^{-2}$}. This estimate,
however, does not reproduce the $m_\pi$-dependence of our
nucleon mass data. Using \mbox{$2 d_{16}^{r} -
  d_{18}^{r}=3.0\,\textrm{GeV}^{-2}$} instead results in much better fits to
the data. The $O(m^5_\pi)$ curves in
Fig.~\ref{fig:ComparingDiffOrders} have been calculated with both these
numbers. 

The corresponding expansion of the $\sigma$ term is calculated 
from the derivative of $M_{N}$ with respect to the light quark 
mass according to Eq.(\ref{eq:sigmafh}). Expressed in terms of
$m_{\pi}$ it reads
\begin{equation}\begin{split}
\sigma =
\tilde{h}_{1}m_{\pi}^{2}+\tilde{h}_{2}m_{\pi}^{3}+\tilde{h}_{3}&m_{\pi}^{4}
\log\left(\frac{m_{\pi}}{\lambda}\right)+\tilde{h}_{4}m_{\pi}^{4}\\
&+\tilde{h}_{5
} m_{\pi}^{5}\log\left(\frac{m_{\pi}}{\lambda}\right)+\tilde{h}_{6}m_{\pi}^{5}
+\mathcal{O}(m_{\pi}^{6})\,,
\end{split}
\end{equation}
with the coefficients
\begin{eqnarray*}
\tilde{h}_{1} &=& -4c_{1}\,,\qquad\qquad
\tilde{h}_{2} = -\frac{9(g^0_A)^{2}}{64\pi (F^0_\pi)^{2}}\,,\\
\tilde{h}_{3} &=& -\frac{3(g^0_A)^{2}-28M_{0}c_{1}+3M_{0}c_{2}+12M_{0}c_{3}}
                        {16M_{0}\pi^{2}(F^0_\pi)^{2}}\,,\\
\tilde{h}_{4} &=& 8e_{1}^{r}-\frac{8c_{1}l_{3}^{r}}{(F^0_\pi)^{2}}
 -\frac{3}{64M_{0}\pi^{2}(F^0_\pi)^{2}}
          \left(3(g^0_A)^{2}-8M_{0}c_{1}+4M_{0}c_{3}\right)\,,\\
\tilde{h}_{5} &=& \frac{3(g^0_A)^{2}(40(g^0_A)^{2}-3)}{1024\pi^{3}(F^0_\pi)^{4}}\,,\\
\tilde{h}_{6} &=&
\frac{3g^0_A}{2048M_{0}^{2}\pi^{3}(F^0_\pi)^{4}}
    \left \{ 3g^0_A(12(g^0_A)^{2}-1)M_{0}^{2}\right.\\
  &&\left. {} +4\pi^{2}
\left[ 16g^0_A(5l_{4}^{r}-3l_{3}^{r})M_{0}^{2}
      +5(F^0_\pi)^{2}(g^0_A-32M_{0}^{2}(2d_{16}^{r}
-d_{18}^{r}))\right]\right \} \,.
\end{eqnarray*}

Finally, we add a remark concerning the LEC $e^r_1$. Becher and
Leutwyler give an estimate for the Delta contribution to
$\bar{e}_{1}^{BL}$ (see Appendix~D of \cite{Becher:1999he}):
\begin{equation}
\bar{e}_{1,\Delta}^{BL} =
 -\frac{g_{\Delta}^{2}}{48\pi^{2}\Delta}
  \left(6\log\left(\frac{\Delta}{2M_{0}}\right) + 5\right)\,. 
\end{equation}
For $g_{\Delta}=13.0\,\mathrm{GeV}^{-1}$,
$\Delta=(1.232-0.939)\,\mathrm{GeV}$, this amounts to
$\bar{e}_{1,\Delta}^{BL}\sim 7.5\,\mathrm{GeV}^{-3}$. Translated to
our choice of LECs, this would give 
\begin{align}\nonumber
  4\tilde{e}_{1}^{r}(\lambda=M_{N}) 
  &= 4\bar{e}_{1} 
  +\frac{3(g^{2}-8c_{1}M_{0}+c_{2}M_{0}+4c_{3}M_{0})}{32M_{0}\pi^{2}(F^0_\pi)^{2}}
  \log\left(\frac{m_{\mathrm{phys}}}{M^{\mathrm{phys}}_{N}}\right)
  = 7.5\,\mathrm{GeV}^{-3}\\
 &\Rightarrow -1\,\mathrm{GeV}^{-3} \,\lesssim\,
 e_{1}^{r}(\lambda=0.138\,\mathrm{GeV})\,\lesssim\, 1\,\mathrm{GeV}^{-3}\,, 
\label{eq:e1r_range}
\end{align}
so this (shifted) LEC may therefore be expected to be of natural
size. Such resonance saturation estimates usually give only very rough
estimates for the higher-order LECs, at least in the baryonic
sector. Nonetheless, such a range of values for $e^r_1$ would be
consistent with the findings of our fits (cf.~\ref{app:fitdetails}).

\section{More details on the fits}
\label{app:fitdetails}

\begin{figure*}
  \centering
  \mbox{\includegraphics[height=5.3cm]{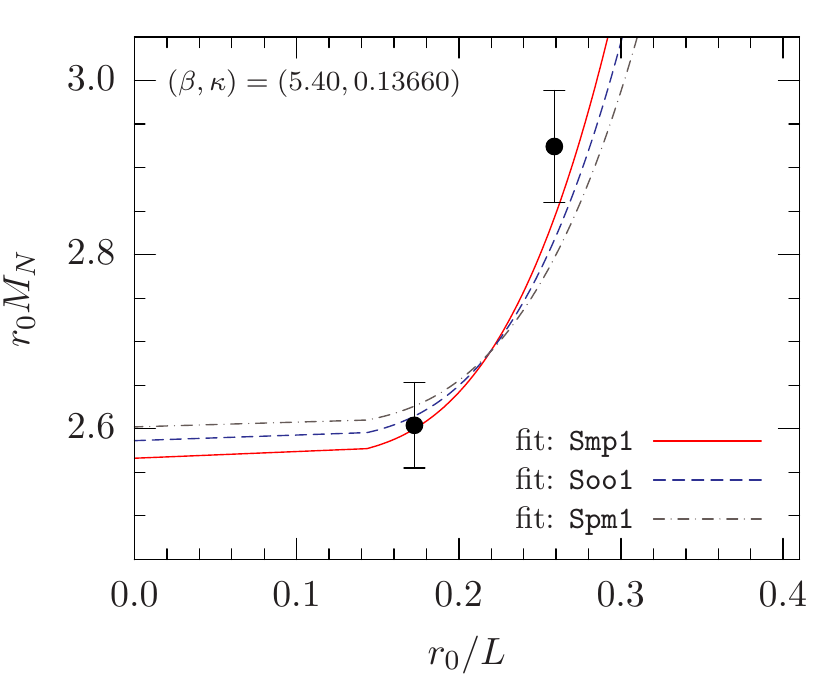}
  \includegraphics[height=5.3cm]{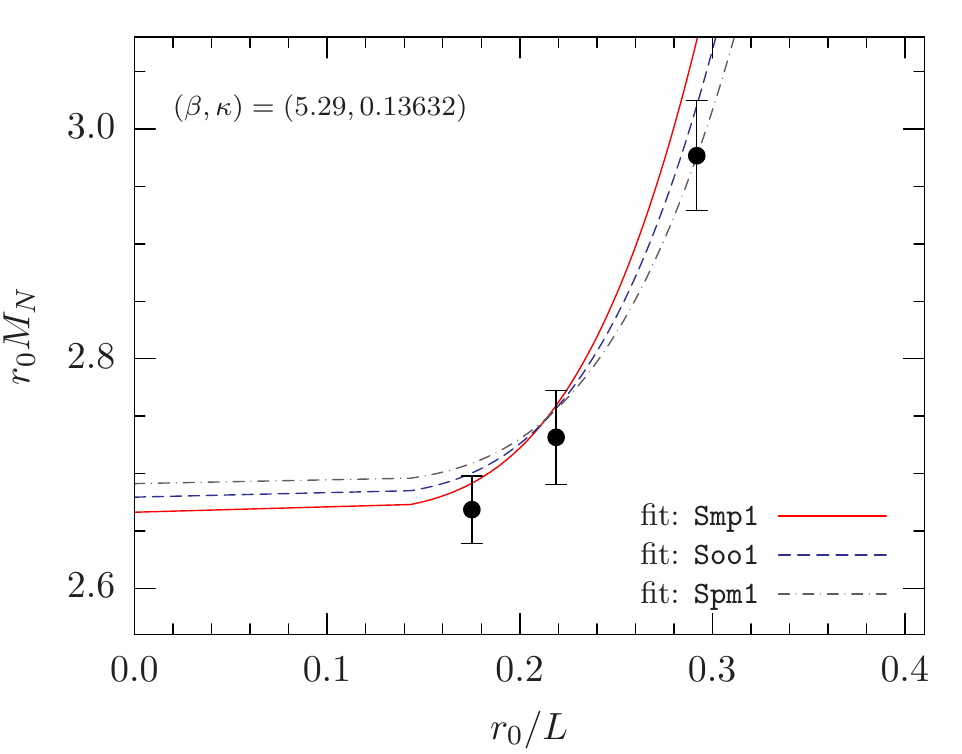}}
  \caption{Volume dependence of our nucleon mass data at
$(\beta,\kappa)=(5.29,0.13632)$ (left) and $(5.40, 0.13660)$ (right)
resulting from different combined fits, labeled \texttt{Smp1},
\texttt{Soo1} and \texttt{Spm1} in \Tab{tab:fitparaS}.  They
illustrate the maximum variation of the finite-volume corrections for
all our combined fits [for \mbox{$(r_0m_\pi)^2<1.3$}] with the parameters
$c_3$ and $\bar{l}_3$.}
  \label{fig:Mn_finV}
\end{figure*}

Here we give a short summary and discussion of the fit parameters for
our (simultaneous) fits to the nucleon mass (and $\sigma$-term)
data. For both types, fits have been performed for different fit
ranges and fixed input parameters, $c_2$, $c_3$ and $\bar{l}_3$, in
order to systematically explore the dependence of the fit parameters
on this external bias.  \Tab{tab:fitparaS} summarizes the results for
the combined fits and \Tab{tab:fitparaN} those for the stand-alone
fits.

\begin{figure*}
 \begin{center}
 \mbox{\includegraphics[width=0.4\linewidth]{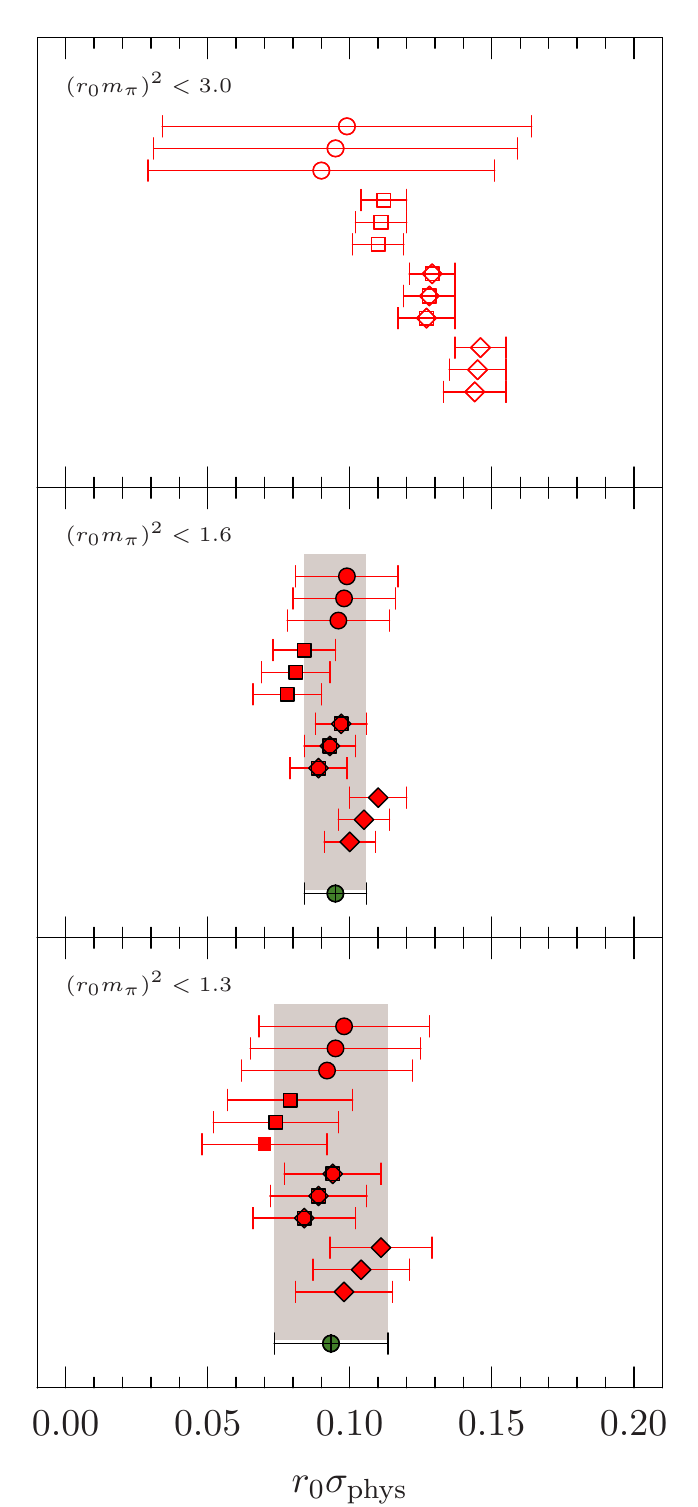}
 \quad\includegraphics[width=0.4\linewidth]{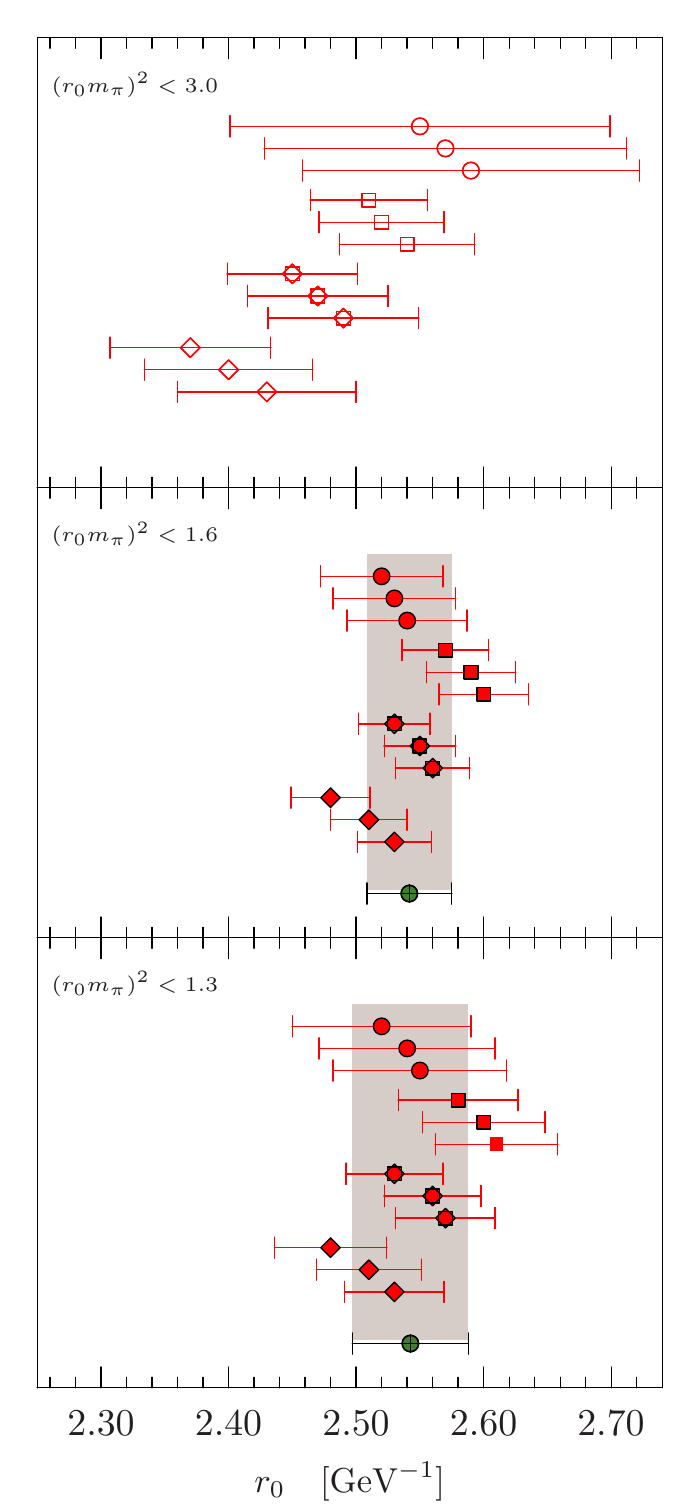}}
 \end{center}
\caption{Values for $r_0\sigma$ at the physical point (left) and $r_0$
    (right) from the combined fits listed in \Tab{tab:fitparaS}. 
    From top to bottom, panels are ordered according to a
decreasing upper limit for $(r_0m_\pi)^2$. Within each panel, data
points are grouped with respect to the values chosen for $c_{3}$ and
$\bar{l}_3$:  Different (red) symbols refer to different values for $c_{3}$
(circles: $c_3=\mbox{free}$; squares: $c_3=-4.7$+1.3; polygons:
$c_3=-4.7$; diamonds: $c_3=-4.7$-1.3, all in GeV$^{-1}$), while triples
of neighboring points (same symbol) refer to different values
of $\bar{l}_3=3.2$+0.8, 3.2, 3.2-0.8 (from top to bottom).
Open symbols refer to poor fits ($\chi^2_r>2.0$), full symbols to good fits
($\chi^2_r<2.0$) and black-framed full symbols to fits where $\chi^2_r<1.3$.
If applicable, gray error bands are shown for weighted averages (green
symbol at the bottom of each band) of the black-framed points.}
\label{fig:r0sigma_r0_r0mpi2_max}
\end{figure*}

\begin{table*}
 \begin{center}\scriptsize
\begin{tabular}{@{\hspace{-3em}}r@{\hspace{1em}}c@{\,}rll@{\;\;\;}l@{\quad}l@{
\quad}l@{ \hspace {
0.5em}}ll}
  \hline\hline
  \multicolumn{1}{l}{\rule{0ex}{4ex} Key} & $(r_0m_\pi)_{\max}^2$ &
\multicolumn{1}{c}{$\chi^2_r$}&
$r_0M_0$ &
$c_1/r_0$ & $e_1^r/r_0^3$ & $c_2/r_0$ & $c_3/r_0$  & $r_0\sigma_{\mathrm{\phys}}$
&\hspace{-1em}$r_0\,\left[1/\mathrm{GeV}\right]$ \\
  \hline 
$\star$ \tt Smm1 & 1.30 & 4.470 / 5 & 2.26(4) & -0.34(3) & -0.142(11) & 1.30 & -2.37 & 0.098(17) & 2.53(4) \\
$\star$ \tt Smo1 & 1.30 & 4.450 / 5 & 2.22(4) & -0.36(3) & -0.148(10) & 1.32 & -2.39 & 0.104(17) & 2.51(4) \\
$\star$ \tt Smp1 & 1.30 & 4.810 / 5 & 2.19(4) & -0.39(3) & -0.155(10) & 1.33 & -2.42 & 0.111(18) & 2.48(4) \\
$\star$ \tt Som1 & 1.30 & 4.550 / 5 & 2.31(4) & -0.28(3) & -0.022(11) & 1.28 & -1.83 & 0.084(18) & 2.57(4) \\
$\star$ \tt Soo1 & 1.30 & 4.060 / 5 & 2.29(4) & -0.29(3) & -0.024(11) & 1.29 & -1.84 & 0.089(17) & 2.56(4) \\
$\star$ \tt Sop1 & 1.30 & 3.690 / 5 & 2.26(4) & -0.31(3) & -0.027(10) & 1.30 & -1.85 & 0.094(17) & 2.53(4) \\
\tt Spm1 & 1.30 & 6.990 / 5 & 2.37(4) & -0.22(3) & 0.087(12) & 1.26 & -1.30 & 0.070(22) & 2.61(5) \\
$\star$ \tt Spo1 & 1.30 & 6.380 / 5 & 2.35(4) & -0.23(3) & 0.087(11) & 1.27 & -1.31 & 0.074(22) & 2.60(5) \\
$\star$ \tt Spp1 & 1.30 & 5.770 / 5 & 2.33(4) & -0.25(3) & 0.087(11) &
1.28 & -1.32 & 0.079(22) & 2.58(5) \\
\hline
ave. & 1.30 & $\chi^2_r<1.3$      & 2.28(4) & -0.31(3) & -0.048(11)  & ---
 & --- & 0.093(18) & 2.54(4) \\*[1ex]
$\star$ \tt Sfm1 & 1.30 & 4.180 / 4 & 2.28(7) & -0.31(7) & -0.089(127) & 1.29 & -2.13(58) & 0.092(30) & 2.55(7) \\
$\star$ \tt Sfo1 & 1.30 & 3.850 / 4 & 2.26(7) & -0.32(7) & -0.073(125) & 1.30 & -2.06(56) & 0.095(30) & 2.54(7) \\
$\star$ \tt Sfp1 & 1.30 & 3.620 / 4 & 2.25(8) & -0.33(7) & -0.054(122)
& 1.31 & -1.98(55) & 0.098(30) & 2.52(7) \\
\hline
 ave. & 1.30 & $\chi^2_r<1.3$          & 2.26(7)  & -0.32(7)  & -0.072(125) &
--- & -2.05(56) & 0.095(3) & 2.54(7) \\*[1ex]

\tt \\
$\star$ \tt Smm2 & 1.60 & 5.140 / 7 & 2.25(3) & -0.35(2) & -0.146(5) & 1.30 & -2.37 & 0.100(9) & 2.53(3) \\
$\star$ \tt Smo2 & 1.60 & 5.100 / 7 & 2.22(3) & -0.36(2) & -0.150(5) & 1.32 & -2.39 & 0.105(9) & 2.51(3) \\
$\star$ \tt Smp2 & 1.60 & 5.460 / 7 & 2.19(4) & -0.38(2) & -0.154(5) & 1.33 & -2.42 & 0.110(10) & 2.48(3) \\
$\star$ \tt Som2 & 1.60 & 5.270 / 7 & 2.30(3) & -0.29(2) & -0.028(5) & 1.29 & -1.83 & 0.089(10) & 2.56(3) \\
$\star$ \tt Soo2 & 1.60 & 4.730 / 7 & 2.28(3) & -0.31(2) & -0.029(5) & 1.30 & -1.84 & 0.093(9) & 2.55(3) \\
$\star$ \tt Sop2 & 1.60 & 4.350 / 7 & 2.25(4) & -0.32(2) & -0.030(5) & 1.30 & -1.86 & 0.097(9) & 2.53(3) \\
$\star$ \tt Spm2 & 1.60 & 7.820 / 7 & 2.35(3) & -0.24(2) & 0.079(5) & 1.27 & -1.31 & 0.078(12) & 2.60(4) \\
$\star$ \tt Spo2 & 1.60 & 7.100 / 7 & 2.33(3) & -0.25(2) & 0.080(5) & 1.28 & -1.32 & 0.081(12) & 2.59(4) \\
$\star$ \tt Spp2 & 1.60 & 6.430 / 7 & 2.31(4) & -0.26(2) & 0.081(5) &
1.28 & -1.32 & 0.084(11) & 2.57(3) \\
\hline
 ave. & 1.60 & $\chi^2_r<1.3$      & 2.28(4)  & -0.31(2) & -0.033(5) & --- &
--- & 0.095(10) & 2.54(3) \\*[1ex]
$\star$ \tt Sfm2 & 1.60 & 4.870 / 6 & 2.27(6) & -0.32(6) & -0.095(123) & 1.30 & -2.15(57) & 0.096(18) & 2.54(5) \\
$\star$ \tt Sfo2 & 1.60 & 4.510 / 6 & 2.26(6) & -0.33(6) & -0.077(121) & 1.30 & -2.07(55) & 0.098(18) & 2.53(5) \\
$\star$ \tt Sfp2 & 1.60 & 4.280 / 6 & 2.24(6) & -0.33(6) & -0.055(118) & 1.31 & -1.97(54) & 0.099(18) & 2.52(5) \\\hline 
     ave. & 1.60 & $\chi^2_r<1.3$  & 2.26(6) & -0.33(6) & -0.08(12) & --- &
-2.1(6) & 0.098(18) & 2.53(5) \\*[1ex]
\tt \\
\tt Smm3 & 3.00 & 50.650 / 9 & 2.11(3) & -0.48(1) & -0.178(5) & 1.36 & -2.47 & 0.144(11) & 2.43(7) \\
\tt Smo3 & 3.00 & 44.340 / 9 & 2.08(3) & -0.49(1) & -0.180(5) & 1.38 & -2.50 & 0.145(10) & 2.40(7) \\
\tt Smp3 & 3.00 & 38.890 / 9 & 2.05(3) & -0.51(1) & -0.182(6) & 1.39 & -2.53 & 0.146(9) & 2.37(6) \\
\tt Som3 & 3.00 & 36.360 / 9 & 2.18(3) & -0.40(1) & -0.045(5) & 1.33 & -1.89 & 0.127(10) & 2.49(6) \\
\tt Soo3 & 3.00 & 31.150 / 9 & 2.16(3) & -0.40(1) & -0.043(5) & 1.34 & -1.91 & 0.128(9) & 2.47(6) \\
\tt Sop3 & 3.00 & 26.420 / 9 & 2.14(3) & -0.41(1) & -0.041(5) & 1.35 & -1.92 & 0.129(8) & 2.45(5) \\
\tt Spm3 & 3.00 & 29.170 / 9 & 2.25(3) & -0.32(1) & 0.072(4) & 1.30 & -1.34 & 0.110(9) & 2.54(5) \\
\tt Spo3 & 3.00 & 25.220 / 9 & 2.24(3) & -0.33(1) & 0.075(4) & 1.31 & -1.35 & 0.111(9) & 2.52(5) \\
\tt Spp3 & 3.00 & 21.550 / 9 & 2.23(3) & -0.34(1) & 0.078(5) & 1.31 & -1.35 & 0.112(8) & 2.51(5) \\
\tt Sfm3 & 3.00 & 27.120 / 8 & 2.33(9) & -0.24(9) & 0.203(159) & 1.27 & -0.68(75) & 0.090(61) & 2.59(13) \\
\tt Sfo3 & 3.00 & 24.190 / 8 & 2.31(10) & -0.26(10) & 0.180(175) & 1.28 & -0.82(82) & 0.095(64) & 2.57(14) \\
\tt Sfp3 & 3.00 & 21.180 / 8 & 2.28(12) & -0.28(11) & 0.159(188) & 1.29 & -0.95(87) & 0.099(65) & 2.55(15) \\
\hline\hline
 \end{tabular}
   \end{center}
\caption{Parameters from our combined fits to the nucleon mass and $\sigma$-term
  data.   Keys in the first column indicate the features of the fits. All keys
start with \texttt{S..} indicating that the $\sigma$-term data has been
included when fitting. Entries are ordered according to increasing           
  $(r_0m_\pi)^2_{\max}$, as specified in column 2. The
$\chi^2$-value together with the number of degrees of freedom
  ($ndf$) is given in column~3. 
  Columns~4 to 8 list the fit
  parameters $M_0$, $c_1$, $e_1^r(0.138\mathrm{MeV})$, $c_2$ and $c_3$
in units of $r_0$. 
  The corresponding estimates for $r_0\sigma$ at the
  physical point and for $r_0$ are given in
  column~9 and 10. Since $r_0$ is not a fit parameter, but
  has been iteratively fixed as explained in the text, the error for $r_0$ is
  that of $\widehat{M}_N(r_0\cdot m^{\phys}_\pi)/M_N^{\phys}$.
  Lines tagged with a $\star$ indicate
  fits where $\chi^2/ndf<1.3$. Parameters of these fits enter the
  weighted averages given in the lines starting with ``ave.``.}
 \label{tab:fitparaS}
\end{table*} 

\begin{table*} 
\begin{center}\scriptsize
\begin{tabular}{@{\hspace{-3em}}r@{\hspace{1em}}c@{\,}rll@{\;\;\;}l@{\quad}l@{
\quad}l@{\hspace{
0.5em}}ll}
  \hline\hline
  \multicolumn{1}{l}{\rule{0ex}{4ex} Key} & $(r_0m_\pi)_{\max}^2$ &
\multicolumn{1}{c}{$\chi^2_r$} &
$r_0M_0$ &
$c_1/r_0$ & $e_1^r/r_0^3$ & $c_2/r_0$ & $c_3/r_0$  & $r_0\sigma_{\mathrm{\phys}}$
&\hspace{-1em}$r_0\,\left[1/\mathrm{GeV}\right]$ \\
  \hline
\tt Nmo1 & 1.30 & 4.760 / 4 & 2.19(9) & -0.39(6) & -0.155(11) & 1.33 & -2.42 & 0.111(40) & 2.48(10) \\
\tt Noo1 & 1.30 & 3.510 / 4 & 2.23(9) & -0.34(6) & -0.030(11) & 1.32 & -1.88 & 0.103(35) & 2.50(9) \\
\tt Npo1 & 1.30 & 4.450 / 4 & 2.26(9) & -0.29(6) & 0.088(11) & 1.31 & -1.35 & 0.096(40) & 2.53(10) \\\hline 
     ave. & 1.30 & $\chi^2_r<1.3$  & 2.23(9) & -0.34(6) & -0.032(11) & --- &
--- & 0.10(4) & 2.50(10) \\*[1ex]
\tt Nfo1 & 1.30 & 3.510 / 3 & 2.23(10) & -0.34(8) & -0.024(133) & 1.32 & -1.85(58) & 0.103(50) & 2.51(12) \\*[1ex]
\tt Nmo2 & 1.60 & 5.280 / 6 & 2.20(7) & -0.38(4) & -0.153(8) & 1.32 & -2.41 & 0.109(23) & 2.49(7) \\
\tt Noo2 & 1.60 & 4.200 / 6 & 2.23(7) & -0.34(4) & -0.031(8) & 1.32 & -1.87 & 0.103(21) & 2.51(6) \\
\tt Npo2 & 1.60 & 5.170 / 6 & 2.26(7) & -0.30(4) & 0.086(7) & 1.31 & -1.35 & 0.097(24) & 2.52(7) \\\hline 
     ave. & 1.60 & $\chi^2_r<1.3$  & 2.23(7) & -0.34(4) & -0.022(8) & --- &
--- & 0.103(22) & 2.51(6) \\*[1ex]
\tt Nfo2 & 1.60 & 4.200 / 5 & 2.23(8) & -0.34(6) & -0.029(132) & 1.32 & -1.87(58) & 0.103(26) & 2.51(7) \\*[1ex]
\tt Nmo3 & 3.00 & 35.320 / 8 & 1.92(4) & -0.57(1) & -0.191(9) & 1.47 & -2.68 & 0.156(11) & 2.24(8) \\
\tt Noo3 & 3.00 & 17.010 / 8 & 2.04(4) & -0.46(1) & -0.030(8) & 1.41 & -2.00 & 0.138(9) & 2.35(6) \\
\tt Npo3 & 3.00 & 11.140 / 8 & 2.14(4) & -0.37(1) & 0.099(7) & 1.36 & -1.40 & 0.121(8) & 2.43(4) \\
\tt Nfo3 & 3.00 & 10.810 / 7 & 2.18(11) & -0.34(10) & 0.150(156) & 1.34 & -1.15(69) & 0.114(43) & 2.46(12) \\ 
 \hline\hline
 \end{tabular}
   \end{center}
   \caption{Parameters from our fits to the nucleon mass alone. All keys
   start with \texttt{N..} indicating that \underline{no} $\sigma$-term data
   has been included. The remaining character sequence has
   the same meaning as in \Tab{tab:fitparaS}.}
  \label{tab:fitparaN}
\end{table*}

The two tables have to be read as follows: Each line is the result for one
particular fit. In the first column, a unique key is
assigned to each fit. Keys starting with
\texttt{N} (``no $\sigma$``) refer to stand-alone fits to the nucleon mass
data [i.e., \Eq{eq:chi2Mn}], while keys starting with  \texttt{S} (``with
$\sigma$'') signify combined fits [\Eq{eq:chi2Mnsigma}]. The second
and third characters are either \texttt{m}, \texttt{o}, \texttt{p} or
\texttt{f}, depending on the values assigned to the parameters $c_3$ and
$\bar{l}_3$ (see \Sec{sec:parameters}). If the second (third)
character is an~\texttt{o}, $c_3$ ($\bar{l}_3$) was fixed to
$c_3=-4.7$\,GeV$^{-1}$ ($\bar{l}_3= 3.2$); if instead this character is
\texttt{p} (\texttt{m}), $c_3$ and $\bar{l}_3$ were fixed to these values,
\underline{p}lus (\underline{m}inus) one
standard deviation. If the second character is \texttt{f}, $c_3$ was not fixed
but left as a free fit parameter. Since
$c_2$ is known to a much better precision than $c_3$ [see \Eq{eq:c2c3_Meissner}]
it is not varied, but always set to its phenomenological value
$c_2=3.3$~GeV$^{-1}$. Each key ends with an integer
labeling, in increasing order, the upper limit
$(r_0m_\pi)^2_{\max}$ of the fit interval
\begin{equation}
  (r_0m_\pi)^2<(r_0m_\pi)^2_{\max} \,,
\end{equation}
where $(r_0m_\pi)^2_{\max}$ is either $1.3$, $1.6$ or $3.0$ (given in the
second column). In all cases we require $L/r_0 > 3$. To reduce the
finite-volume effect in the pion mass, we exclude data points for all
those ($\beta,\kappa$) combinations for which there is not at least
one value for $r_0m_\pi$ available that satisfies $m_\pi L>3.5$.

All (fixed and floating) fit parameters listed in
Tables \ref{tab:fitparaS} and \ref{tab:fitparaN} are given in units
of~$r_0$. The 
corresponding self-consistent physical value of $r_0$---reached
iteratively for each fit (see \Eq{eq:r0iteration})---is given in the last
column in GeV$^{-1}$. That is, if one is interested in the physical value for a
particular fit parameter, this parameter has to be multiplied by the
corresponding power of $r_0$ given at the end of the same line. In
\Tab{tab:fitparaS} we also give weighted averages of the fit parameters for
fits with a $\chi_r^2<1.3$ (see lines starting with ``ave.``).

Let us now comment our fits.  The fact that our fits with
$(r_0m_\pi)^2<1.6$ yield (in the majority of cases) $\chi^2_r$-values
around one indicates that the constraint $L/r_0>3.0$ on the spatial
lattice extension has been sufficient to correct for the finite-volume
effect in the data. This can be seen also from \Fig{fig:Mn_finV},
where we compare data for $r_0M_N$ at fixed $(\beta,\kappa)$ but
different $L/r_0$ to the fitted function at the corresponding
$r_0m_\pi$.  If we had relaxed the constraint $L/r_0>3$, the
finite-volume effect could not be completely compensated for data
points with $L/r_0\le3$.  This is, for example, the case for our
points at $(\beta,\kappa)=(5.40,0.13640)$ from a $24^3\times 48$
lattice.

Looking at the fit parameter $e^r_1$, our fits are less robust, however. In
fact, the sign of $e^r_1$ is strongly correlated with the value we choose for
$c_3$. If we set, for example, $c_3=-4.7+1.3\,\textrm{GeV}^{-1}$ all fits result
in a positive $e^r_1$, while, if we set $c_3=-4.7\,\textrm{GeV}^{-1}$ or
$c_3=-4.7-1.3\,\textrm{GeV}^{-1}$, $e^r_1$ comes always out
negative.  The value we choose
for $\bar{l}_3$ does not affect the sign of $e^r_1$, yet $\bar{l}_3$ has a minor
effect on $e^r_1$'s absolute value. To our knowledge, nothing is really known
about the sign of $e^r_1$, so we do not restrict it. From our data we can say
it tends to be negative if $c_3\le-4.7\textrm{GeV}^{-1}$ and positive if
$c_3\ge-3.4\,\textrm{GeV}^{-1}$. If we leave $c_3$ as free fit parameter,
we obtain values around $-5.0\,\mathrm{GeV}^{-1}$ for
$c_3$ ($\pm 1.5\textrm{GeV}^{-1}$ statistical uncertainty), and
$e^1_r$ is consistent with zero within errors. This is found for fits
where $(r_0m_\pi)^2_{\max}$ is either 1.6 or 1.3. Interestingly, the
range of fitted values for $e^r_1$ lies in the ballpark expected from
BChPT (see \Eq{eq:e1r_range}).

A correlation with $c_3$ is also seen for $r_0\sigma_{\mathrm{phys}}$ and $r_0$ (see
columns 9 and 10 in Tables \ref{tab:fitparaS} and
\ref{tab:fitparaN}). For larger, i.e., less negative values of $c_3$,
the results for $r_0\sigma_{\mathrm{phys}}$ tend to smaller numbers,
while $r_0$ tends to larger values.

For the reader's convenience, we visualize the variation of
the values for $r_0$ and $r_0\sigma_{\mathrm{\phys}}$ in
\Fig{fig:r0sigma_r0_r0mpi2_max} (only for results from
\Tab{tab:fitparaS}). From top to bottom, panels are ordered
with decreasing $(r_0m_\pi)^2_{\max}$, while within each panel, points
are grouped according to the values chosen for $c_3$ and
$\bar{l}_3$. Symbols distinguish different $c_3$, neighboring points
with the same symbol are for different $\bar{l}_3$. The color
intensity of each point is related to the $\chi^2_r$-value of each
point. When applicable each panel also shows the corresponding weighted
average.


\bibliographystyle{apsrev}

\end{document}